\documentclass[%
superscriptaddress,
 amsmath,
 amssymb,
 aps,
prb,
floatfix,
twocolumn
]{revtex4-2}
\usepackage[utf8]{inputenc}
\usepackage[T1]{fontenc}
\usepackage{babel}
\usepackage{graphicx}
\usepackage{amsfonts}
\usepackage{amssymb}
\usepackage{amsmath}
\usepackage{amsthm}
\usepackage{mathtools}
\usepackage{physics}
\usepackage[normalem]{ulem}
\usepackage{blindtext}
\usepackage{dsfont}

\def\id{{\rm 1\kern-.22em l}}

\usepackage{mathtools}
\usepackage{braket}
\renewcommand\bra[1]{{\langle{#1}|}}
\renewcommand\ket[1]{{|{#1}\rangle}}
\usepackage{graphicx}
\usepackage[format=plain,justification=centerlast]{caption}
\usepackage[format=plain,justification=centerlast]{subcaption}
\usepackage{bm}
\usepackage[dvipsnames]{xcolor}

\usepackage{natbib}
\setcitestyle{square,numbers}
\usepackage[colorlinks]{hyperref}

\begin{document}

\title{Superresolution in separation estimation between two dynamic incoherent sources using spatial demultiplexing} 

\author{Konrad Schlichtholz}
\email[Corresponding author:]{konrad.schlichtholz@phdstud.ug.edu.pl}
\affiliation{International Centre for Theory of Quantum Technologies (ICTQT),
University of Gdansk, 80-308 Gdansk, Poland}
\author{Łukasz Rudnicki}
\affiliation{International Centre for Theory of Quantum Technologies (ICTQT),
University of Gdansk, 80-308 Gdansk, Poland}
\affiliation{Center for Photonics Sciences, University of Eastern Finland, P.O. Box 111, FI-80101 Joensuu, Finland}
\begin{abstract}
Achieving resolution in the sub-Rayleigh regime (superresolution) is one of the rapidly developing topics in quantum optics and metrology.  Recently, it was shown that perfect measurement based on spatial mode demultiplexing (SPADE) in Hermite-Gauss modes allows one to reach the quantum limit of precision for estimation of separation between two weak incoherent stationary sources. Since then, different imperfections such as misalignment or crosstalk between modes have been studied to check how this result translates into more realistic experimental setups. In this paper, we consider another deviation from the perfect setup by discarding the assumption about the stationarity of the sources. This is relevant for example for astrophysical applications where planets necessarily orbit around the star. We analyze two examples of dynamics: rotations and oscillations, showing the robustness of the SPADE-based measurement against them. The analysis is based on Fisher information, which allows one to obtain the precision limit through  Cram\'er-Rao bound. Furthermore, we formulate a measurement algorithm that allows for the reduction of one parameter for estimation (system orientation angle) in
the stationary sources scenario, maintaining the measurement
precision despite the lack of knowledge about this parameter.
\end{abstract}

\maketitle
\section{Introduction}
The task of resolving the separations between two point-like light sources is one of the core goals of modern optics and metrology.
Traditional measurement approaches based on direct imaging suffer a loss of efficiency with decreasing separations when they are below the so-called Rayleigh regime. This regime is determined by the width of the point spread function of the optical device \cite{resolution_survey_denDekker_1997,Fourier_optics_Goodman_2005,Rayleigh_curse_Paur_2018,Rayleigh_curse_Larson_2018}. Such a phenomenon is referred to as Rayleigh's curse.

A variety of techniques have been developed to overcome the limits of direct imaging and achieve the so-called superresolution.
This comprises, among other methods, optical reconstruction microscopy \cite{STORM_Hell_2007}, superoscillations \cite{superoscillations_Smith_2016,superoscillations_Gbur_2019},  photoactivated localization microscopy \cite{PALM_Betzig_2006,PALM_Hess_2006}, stimulated-emission microscopy \cite{Hell:94,Klar8206}, inversion of coherence along an edge \cite{SPLICE_Tham_2017,SPLICE_Bonsma-Fisher_2019}, and measurements based on homodyning \cite{Hsu_2004,Delaubert:06}. In particular, one of recent methods utilizes spatial demultiplexing (SPADE) in Hermite-Gauss modes. It was shown to achieve the optimal measurement allowed by quantum metrology for estimating the distance between two incoherent
Poissonian light sources \cite{superresolution_Tsang_2016,superresolution_starlight_Tsang_2019}. In recent years, this method has been actively experimentally tested \cite{Santamaria:23,Santamaria:24,Rouviere:24}

However, any measurement scheme cannot be perfectly implemented in an experimental setup. Therefore, imperfections must be included in the analysis to obtain reliable estimates. Until now, different imperfections have been considered, such as noise \cite{Len, metrology_noise_Kolodynski_2013}, apparatus misalignment \cite{superresolution_Tsang_2016}, and crosstalk between measurement modes \cite{Manuel,Giacomo,Linowski}. All of these considerations have resulted in the re-establishment of Rayleigh's curse in the limit of small separations. In fact, all possible imperfections impact the precision of the measurement by modifying the probabilities of the results. Therefore, one can observe that an analogous effect to imperfections of the measurement apparatus might have the motion of the system itself, which in an obvious way modifies the probabilities of outcomes.  Thus, the motion of the sources should be considered as another factor that could impact the precision limits of the measurement and has to be accounted for to avoid errors from mismatches between the model and data.   One of the primary objects of interest that one would like to study using techniques like SPADE are binary stars and exoplanets \cite{hypothesis_testing_exoplanets_2021,Exo_5,Hyp_1, Schlichtholz:24}. This is because this method was shown to be capable of performing with the maximal precision allowed by quantum mechanics in the task of distinguishing between two versus one sources in the stationary scenario \cite{Hyp_1}. However, such types of system necessarily undergo motion on their orbits, which can have an impact on the analysis of long-term observation data.  In particular, the periodic motion of the systems is one of the main features utilized in exoplanet detection with, e.g., Doppler spectroscopy and transit photometry \cite{Exo_te}. Therefore, rotations of the system are of particular interest for such applications. While the basic task is to distinguish these types of systems, in order to perform an optimal hypothesis test for a given imaging method, it is necessary first to estimate the potential separation between the objects in the system \cite{Schlichtholz:24}.

In this paper, we show a method for considering the dynamics of sources in parameter estimation tasks. Based on this, we analyze the impact of dynamics on the estimation of the separation between two weak sources assessed with SPADE. In particular, we analyze the dynamics in the form of rotations, which is relevant for astronomical applications. In particular, we show that the precision of SPADE is robust against the dynamics of the system orientation, and thus such motions do not influence the ability of achieving a superresolution. Further on, we consider another example of estimating the average separation between two oscillating sources. Finally, we show how one can eliminate the need for estimating the orientation of the system on the imaging plane, maintaining measurement precision despite lack of knowledge about this parameter.
\section{Measurement scenario}
We consider the scenario where we try to estimate the distance below Rayleigh's limit between two weak incoherent point sources, the positions of which evolve in time in a non-relativistic manner. Here, the weak stands for the photon detection being approximately a Poissonian process.  In order to achieve that spatial mode demultiplexing in Hermite-Gauss modes $\lbrace u_{n,m}(\vec{r})\rbrace$ is performed on the imaging plane $xy$ which is followed by perfect photon counting in the modes. This measurement is performed for some finite time $\tau_M$. However, the time of specific counts is not measured. 

We consider the following additional assumptions. As the process is Poissonian and we can discard the no-count events, we can effectively describe the measurement as measurement on $N$ single-photon states. The state of each photon before measurement is described by some density matrix $\hat\rho(q_i,p_i,\vec{P})$ characterized by some point in the phase space of the system of the sources. Here we denote generalized positions as $q_i$,  momenta as $ p_i$, and some additional parameters which we denote as $\vec{P}$ (some possibly redundant depending on the choice of $q_i$ like, for example, separation when one chooses $q_i$ to be a position of sources in Cartesian coordinates). Because motion is non-relativistic, we can as an approximation neglect the influence of the momenta, and thus we put:
\begin{equation}
  \hat\rho(q_i,p_i,\vec{P})\approx\hat\rho(q_i,0,\vec{P}):=  \hat\rho(q_i,\vec{P}).
\end{equation}
We further assume that the sources are objects undergoing classical evolution in phase space  (what is clear when one considers stars) and that the motion is known and described by some well-behaved trajectory $q_i(t)$. Furthermore, the measurement time is much longer than the average time between photon detections $\tau_d$, i.e., $\tau_M\gg \tau_d$. Lastly, the point spread function of the diffraction-limited measurement apparatus is well described by a Gaussian $u_{00}(\vec{r}-\vec{r}_i)$ \cite{goodman1985} where $u_{00}(\vec{r})=\sqrt{2/(\pi w^2)}\exp{-r^2/w^2}$ and $\vec{r}_i$ describe the position of the $i$ -th source after projection into the $xy$ plane in relation to the origin of the measurement basis. Then, allowing two sources to have different brightnesses, the single-photon state on the imaging plane can be described as the following density matrix:
\begin{equation}
    \rho(q_i,\vec{P})\approx v\ket{\phi_1}\bra{\phi_1}+(1-v)\ket{\phi_2}\bra{\phi_2},
\end{equation}
where $0< v <1$ is a parameter that determines the relative brightness of the sources (where $v=1$ would stand for only the first source visible and $v=0$ for the opposite case) and
\begin{equation}
    \ket{\phi_i}=\int_{\mathbf{R}^2} d\vec{r}\;u_{00}(\vec{r}-\vec{r}_i)\ket{\vec{r}},
\end{equation}
where $\ket{\vec{r}}$ is the single-photon position generalized eigenstate in the imaging plane.

\subsection{Probability of measuring $(n,m)$ Hermite-Gauss mode in stationary case }
One can show that for two stationary incoherent sources  of weak thermal or coherent light probability of measuring photon in $(n,m)$ Hermite-Gauss mode is given by the following \cite{Linowski}:
\begin{equation}
    p(nm|\vec{r}_i,v,d)=v| f_{nm}(\vec{r}_1)|^2+(1-v)| f_{nm}(\vec{r}_2)|^2,\label{probNM}
\end{equation}
where $ f_{nm}(\vec{r}_i)$ are overlap integrals between $(n,m)$ Hermite-Gauss mode $u_{nm}$ and the spatial distribution of the field coming from the $i$-th source. Here, we have chosen as our coordinates $q_i$ the position in the imaging plane $\vec r_i$. These integrals can be put as:
\begin{equation}
 f_{nm}(\vec{r}_i)=\int_{\mathbf{R}^2} d\vec{r}u_{nm}^*(r)u_{00}(\vec{r}-\vec{r}_i).
\end{equation}
In the special case in which the origin of the imaging plane is located exactly in the middle between the sources, the probability (\ref{probNM}) is independent of the relative brightness and is equal to (see Appendix \ref{app:calc} for details):
\begin{equation}\label{eq:probHM_sym}
    p(nm|\vec{r}_i,v,d)=|f_{nm}(\vec{r}_1)|^2.
\end{equation}
Thus, in the following, whenever we consider such a type of system, we will remove $v$ from the list of our relevant parameters.
\section{Precision limit of distance estimation}
In this section, we recall metrological tools based on which we will analyze the precision of distance estimation for dynamical scenarios.
In the following, we focus on the estimation of a single parameter $P_e$. For such a case, the bound on the uncertainty of parameter estimation using an unbiased estimator is determined by the Cram\'er-Rao bound \cite{Kay}:
\begin{equation}
\Delta P_e\geq\frac{1}{\sqrt{N F(P_e)}},\label{eq:prec}
\end{equation}
where $F(P_e)$ is the Fisher information (FI) that comes from every measured photon. Consider the case where among parameters describing our state, there is a single parameter for estimation $P_e$ and other parameters $\vec P_k$ are known, i.e., $\vec P=(P_e,\vec P_k)$. Then in the case of Poissonian photodetection in Hermite-Gauss modes FI for estimation of $P_e$  is given by \cite{Chao:16}:
\begin{multline}\label{eq:FI_definition}
   F(P_e,\vec P_k,q_i) \\= \sum_{n,m=0}^M \frac{1}{p(nm|q_i,P_e,\vec P_k)}
        \left( \frac{\partial}{\partial P_e} p(nm|q_i,P_e,\vec P_k) \right)^2\\
        =\sum_{n,m=0}^M F_{nm}(P_e,\vec P_k,q_i),
\end{multline}
where by $M$ we denote the maximal index of Hermite-Gauss mode distinguished by a particular measurement apparatus, and we explicitly write $P_e$ in the probabilities despite possible redundancy.
Maximization of FI over all physically possible measurements results in quantum FI $F_Q(P_e)$, which establishes the optimal bound for the estimation of the parameter considered \cite{Kay}. In the case of separation estimation, it was shown \cite{superresolution_Tsang_2016} that $F_Q(d)=w^{-2}$. What is more, this bound is achieved in the limit of $M\rightarrow\infty$  by SPADE measurement with Hermite-Gauss modes centered on the origin of the system. In addition, the simplified measurement with $M=1$ approaches the quantum limit when $d/2 w\ll 1$.

\section{Including dynamics}
Let us discuss in this section how one can include the dynamics of the system in the analysis. Due to the dynamics of the sources, each measured photon can be emitted from the system described by different configuration of positions $q_i$. Let us denote the part of the phase space of the system that describes positions as $\Gamma_q$   As we do not measure the time of detection, each emission could occur at any point in $\Gamma_q$. Thus, effectively, we perform sequential measurement on the $N$ copy of the mixed state:
\begin{equation}
    \hat \rho'(\vec{P}) =\int_{\Gamma_q}dq_i\,\textbf{p}(q_i|\vec{P}) \hat\rho(q_i,\vec{P}),\label{eq:density_mat}
\end{equation}
where $\textbf{p}(q_i|\vec{P})$ is some probability density function describing distribution of coordinates $ q_i$ conditioned on parameters $\vec{P}$, therefore, fulfilling $\int_{\Gamma_q}dq_i\,\textbf{p}(q_i|\vec{P})=1$ and integral is taken over all coordinates $q_i$.  From this it follows that given projector $\hat\Pi_k$ for some result $k$, the probability of obtaining $k$ is given:
\begin{equation}
   p(k|\vec{P})= \langle\hat \Pi_k\rangle_{\rho'}=\int_{\Gamma_q}dq_i\,\textbf{p}(q_i|\vec{P}) p(k|q_i,\vec{P}),\label{probPH}
\end{equation}
where $p(k|q_i,\vec{P}):=\tr[\hat\Pi_k\hat\rho(q_i,\vec{P})]$. In our measurement scenario, we use in particular probabilities (\ref{probNM}).

It is now important to relate the abstract probability density $\textbf{p}(q_i|\vec{P})$ with the dynamics of the system. Let us notice that given some trajectory of the system $q_i(t)$ we could approximate $ p(k|\vec{P})$ as the time average over the trajectory:
\begin{equation}
  p(k|\vec{P}) \approx \frac{1}{\tau_M}\int_0^{\tau_M}dt  \, p(k|q_i(t),\vec{P}),\label{timeAvg}
\end{equation}
because there is no privileged point in time for detecting a photon and contribution from each state  $\hat\rho(q_i,\vec{P})$ in such a case goes into the mixture with the weight given by the fraction of the time that system spends near  the given phase-space points. However, in general, 
the probability $ p(k|\vec{P})$ in (\ref{timeAvg}) does not have to be strictly the same as the probability estimated by simply collecting photons over time $t_M$ when the system follows the trajectory $q_i(t)$. That is, the effective phase-space distribution $\textbf{p}(q_i|\vec{P})$ for this measurement could differ from this associated with r.h.s. of (\ref{timeAvg}).  The intuitive reason for that is that the accuracy of such an estimation can highly depend on the density of photon detection in time, as higher density is analogous to using a more dense grid for numerical integration. As we cannot influence this density, it can affect how accurately we can probe the statistics coming from a given phase-space point. This is accompanied by the fact that not all trajectories return close to once-visited phase-space points, which would allow for additional probing from it.  Therefore, the description of such an approach to measurement in terms of (\ref{timeAvg}) in general can work well when the changes of a system and thus also of $p(k|q_i(t),\vec{P})$ are slow in time in relation to the density of photon detection. This can be fulfilled in astronomical scenarios where the movement on the imaging plane can be relatively slow. However, there is a class of trajectories for which such a description is strict.

\subsection{Periodic motion}
Let us consider a special case where the trajectory is periodic in time.  Note that the time between emissions is a continuous random variable, thus any point from phase-space is approached arbitrarily close at times of emission at some point in time. From this and from the ergodic Theorem \cite{Ruelle1979,Moore_Ergodic} follows that in the limit of $\tau_M\rightarrow \infty$  the  time average is equal to the phase-space integral, i. e., equality is strict in (\ref{timeAvg}). What is more, this integral does not depend on the initial conditions. Thus, due to periodicity, the integral can be simplified:
\begin{equation}
 p(k|\vec{P}) = \frac{1}{T}\int_0^{T}dt   \, p(k|q_i(t),\vec{P}),    \label{TimeAvg_per}
\end{equation}
where by $T$ we denote the period of motion.  The intuitive reason for this is the fact that $\tau_M\gg T$, and therefore some results coming from one not full period at the beginning of the measurement, do not significantly impact the entire measurement. Observe that in such a limit, the density of emission in time does not play a role for accuracy as long as  $\tau_d$ is finite.

\section{Dynamic orientation angles}
\begin{figure}[ht]
\centering
\includegraphics[width=0.49\textwidth]{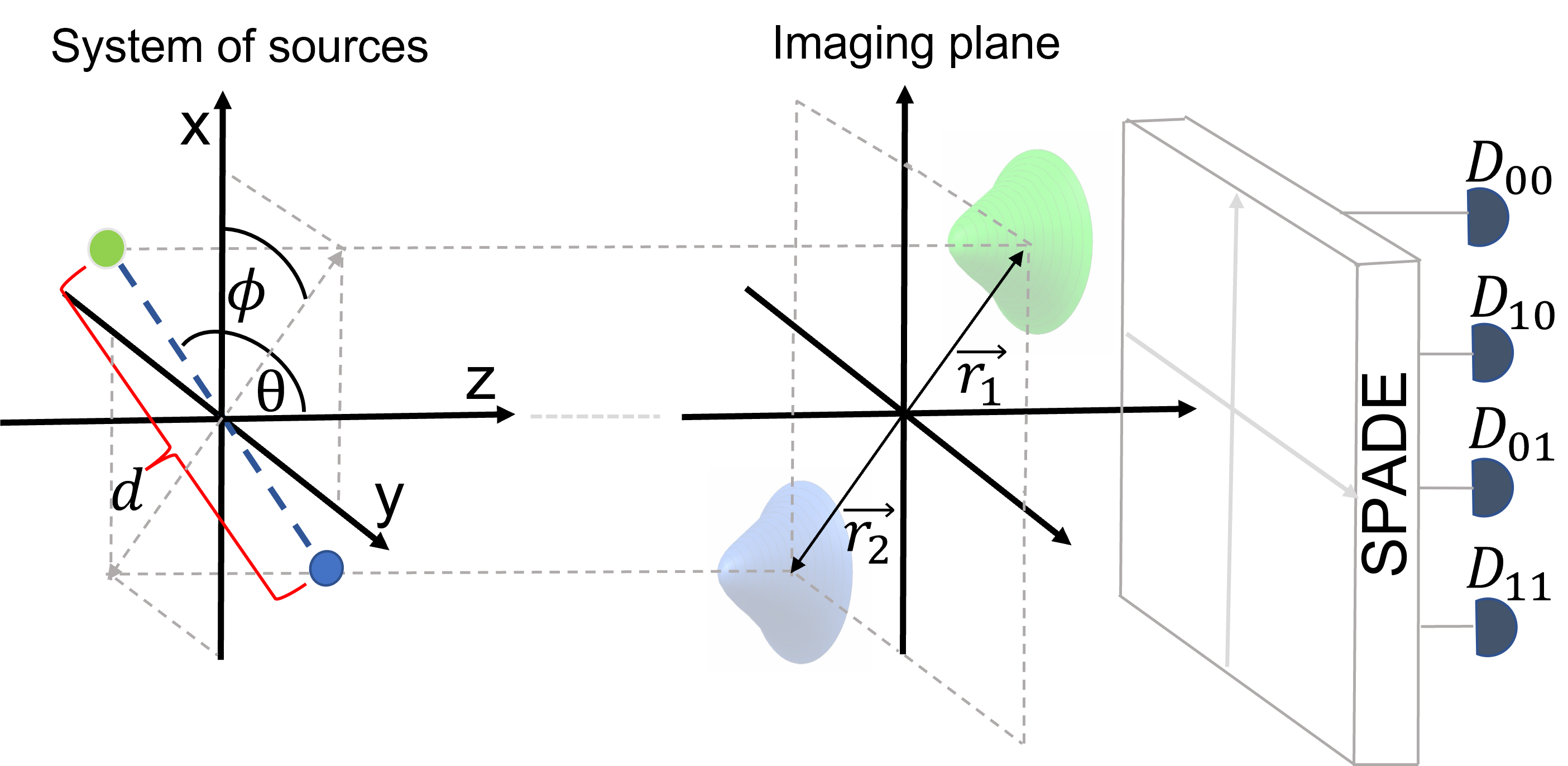} 
\caption{ Schematic representation of the measurement scenario in which the system of two weak incoherent sources separated by $d$ dynamically changes its orientation described by angles $\phi$ and $\theta$. The emitted photons are spatially demultiplexed in the Hermite-Gauss modes on the imaging plane, and the photon counts in the modes $(n,m)$ are registered in detectors $D_{nm}$. The positions of sources in the image plane are described by $\vec{r_i}$.  }
\label{fig:system}
\end{figure}

\begin{figure}
    \centering
    \includegraphics[width=\linewidth]{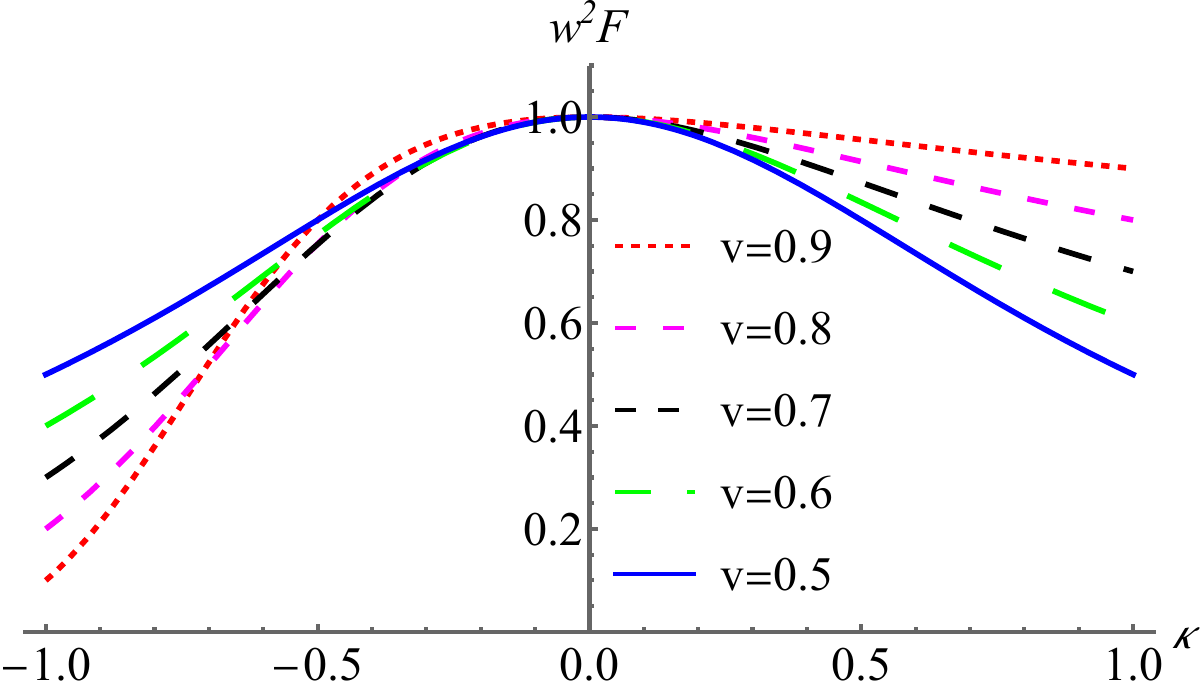}
    \caption{Small separation limit of Fisher information (\ref{eq:FI0}) in the function of $\kappa$ for a system with different relative brightnesses $v$ and with $C=1$ that is, with $\theta=\pi/2$. Note that for relative brightness $v\geq1/2$ part of the figure $\kappa\leq0$  is more relevant for astrophysical scenarios. This is because most commonly one considers a brighter source to have a higher mass, and thus the rotation axis shifted towards it. Although better performance would be obtained if the imaging plane is centered closer to the dimmer source, nonetheless, in all cases the Fisher information is nonzero.}
    \label{fig:kappa}
\end{figure}

Let us analyze an example scenario where we allow a system of two sources with fixed separation between them $d$ to change its orientation in time. However, we restrict possible rotations in the $xy$ plane to rotations with respect to the static axis located between the sources. We assume that the origin of the imaging plane is located on this static axis. To examine the precision achieved in such a system by SPADE, we can compute the FI by replacing the probabilities in (\ref{eq:FI_definition}) with the probabilities (\ref{probPH}) calculated upon (\ref{probNM}). We have two degrees of freedom in the considered system, two orientation angles $(\phi,\theta)$ with $\phi\in [0,2\pi)$ and $\theta\in[0,\pi]$ that describe the position of the first source in spherical coordinates (see Fig. \ref{fig:system}). We choose them as our coordinates $q_i$. However, when the trajectory is considered for simplicity, we do not bound these angles. To fully characterize this system, we additionally need two parameters: the relative brightness $v$ and the position $\xi w\in (-d/2,d/2)$ of the rotation axis with respect to the center of the system of two sources (with $\xi<0$ if the axis is closer to the first source). Then we can write the position of the sources on the imaging plane in the following way: 
\begin{align}
\begin{split}
    \vec{r_1}&=\frac{1}{2}\Big((d+2\xi w) \cos\phi\sin\theta,(d+2\xi w)\sin\phi\sin\theta\Big),\\
    \vec{r_2}&=\frac{1}{2}\Big((d-2\xi w)\cos\phi\sin\theta,(d-2\xi w)\sin\phi\sin\theta\Big).
    \end{split}
\end{align}
Let us state our main result concerning such type of measurement scenario:
\newtheorem{Tw}{Proposition}
\begin{Tw}\label{prop1}
Estimation of the separation between two weak incoherent sources with SPADE in Hermite-Gauss modes (distinguishing at least modes $u_{10},u_{01}$ from others) centered on the rotation axis between sources is free from Rayleigh's curse for any well-behaved distribution of orientation angles (any dynamics of degrees of freedom) for which marginal distribution of angle $\theta$ is not a convex combination of Dirac delta $\delta(\theta)$, and $\delta(\theta-\pi)$. Furthermore, Fisher information $F(d,v,\xi)$ for the estimation of the separation for $d/2w\ll1$ given parameters $v$ and $\xi$ is approximately constant given by the average Fisher information over the orientation distribution:
\begin{equation}
     F(d,v,\xi)\approx\int_0^{2\pi}d\phi\int_0^{\pi}d\theta \,\textbf{p}(\phi,\theta)F(d,\phi,\theta,v,\xi),
\end{equation}
where $\textbf{p}(\phi,\theta)$ is a probability density of angles $\theta$ and $\phi$  $\left(\int d\phi\int d\theta\, \textbf{p}(\phi,\theta)=1\right)$ and $F(d,\phi,\theta,v,\xi)$ is Fisher information for static case with specific orientation of the system. Here equality is strict for $d\rightarrow0$ and also in the special case $\theta(t)=const$ for any $d$.
\end{Tw}
The proof can be found in the Appendix \ref{app:t0}. This result shows that SPADE resembles a particular robustness against this kind of ``noise'' added by the dynamics of the system. This is particularly interesting because even small crosstalk between modes (upon which one in general can model arbitrary passive static imperfections of the system) can result in Rayleigh's curse \cite{Manuel}.  Let us discuss some details concerning Proposition \ref{prop1}. Note that this result shows that FI in such a case is just approximately an average over orientations. In particular, one can find that in the limit $d\rightarrow0$ one has nonzero FI:
\begin{equation}\label{eq:FI0}
\lim_{d\rightarrow0}w^2F(d,v,\xi)=\frac{C(1-\kappa+2\kappa v)^2}{(\kappa-1)^2+4 \kappa v},    
\end{equation}
where $\kappa=2\xi w/d$ and $C$ is coefficient that is dependent only on the marginal distribution $\textbf{p}(\theta)$ of $\theta$ given by
\begin{equation}
    C=\int d\theta\,\textbf{p}(\theta)  \sin^{2} (\theta ) \leq1.
\end{equation}
This coefficient is positive whenever $\textbf{p}(\theta)\neq p_0\delta(\theta)+p_\pi\delta(\theta-\pi) $, where $p_0+p_\pi=1$. This class of distributions corresponds to sources being always aligned along $z$ axis. In such a scenario there is no separation between the sources seen by the measurement apparatus. Thus, in an obvious way for this specific scenario, one obtains no information from the measurement. Let us from now on exclude this pathological case from the analysis.
Observe that Rayleigh's curse is associated with FI decreasing to zero in the limit $d\rightarrow0$. This is because due to inequality (\ref{eq:prec}) in such a case, the measurement becomes harder with decreasing separation. Therefore, for our measurement scenario,  it is easy to see that Rayleigh's curse is not present, as we have in this limit always $F(d,\phi,\theta,v,\xi)>0$. 
Now, assuming $C=1$, one can find that $\lim_{d\rightarrow0}w^2F(d,v,\xi)$ can reach a maximum equal to 1 for $\kappa=0$ and a minimum 0 for $\kappa=1/(1-2v)$. The minimum is always reached for $|\kappa|>1$, but for the system considered $|\kappa|<1$ and therefore one never reaches FI equal to 0. Fig. \ref{fig:kappa} presents the behavior of $\lim_{d\rightarrow0}w^2F(d,v,\xi)$. 
In the following, we will illustrate the impact of the dynamics of the orientation of the system on FI through examples that describe changes of particular orientation angles.

\subsection{$\kappa=0,\,\theta(t)=const$}\label{sec:theta}
\begin{figure}[hb]
\centering
\includegraphics[width=0.49\textwidth]{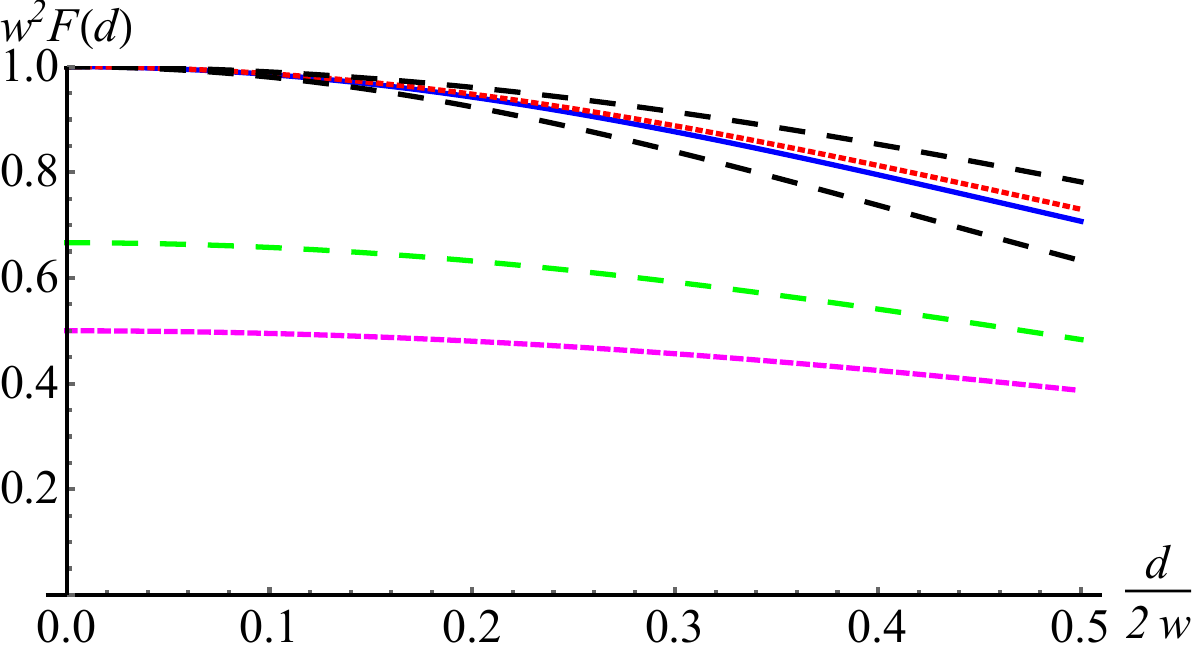} 
\caption{ Fisher information for estimating separation as a function of $d/2w$ for different dynamics of the orientation angles of the system: rotations with constant angular velocity on the $xy$ plane described by $\phi(t)=2\pi t/T$ and $\theta(t)=\pi/2$ (blue curve), oscillations on the $xy$ plane with $\phi(t)=\pi \sin(2\pi t/T)/4$ and $\theta(t)=\pi/2$ (red dotted curve that almost overlaps with the blue curve), rotations with constant angular velocity on $xz$ plane $\theta(t)=2\pi t/T$ and $\phi=0$ (magenta dashed curve), fully random changes of orientation $\textbf{p}(\phi,\theta)=\sin(\theta)/4\pi$ (green long dashed curve). Black long-dashed curves show the stationary case for optimal $\phi=\pi/4$ and the least optimal $\phi=0$ with $\theta=\pi/2$. Note that the Fisher information in the limit $d\rightarrow0$ is nonzero for all cases. This shows that the SPADE-based measurement is robust against the regarded types of motion and that it is free from Rayleigh’s curse.   }
\label{fig:rot}
\end{figure}

Firstly, let us analyze the symmetric scenario with $\xi=0$ and the motion in the $xy$ plane described by $\theta(t)=const$, i.e., with $\mathbf{p}(\phi,\theta)=\mathbf{p}(\phi)\delta(\theta-\theta(0))$.
Based on Proposition \ref{prop1} for periodic trajectories FI in this case can be written as:
\begin{align}
\begin{split}
   \bar  F(d,\theta)=&\int_0^{2\pi}d\phi \,\textbf{p}(\phi)F(d,\phi,\theta)\\=&\frac{1}{T}\int_0^{T}dt \,F(d,\phi(t),\theta)\label{eq:FI_phi}
    \end{split}
\end{align}
with equality being strict. 

Now, we consider two periodic motions. The first type of motion is of particular importance in potential astrophysical applications, as it corresponds, for example, to a system of binary stars of similar mass, that is, rotation with constant angular velocity, which is also equivalent to fully uniformly random orientations in the $xy$ plane. In this case, we have $\textbf{p}(\phi)=1/2\pi$ or $\phi(t)=2\pi t/T$. As another example, we consider the oscillations of $\phi$ in the range $[0,\pi/4]$ with a trajectory given by $\phi(t)=\pi \sin(2\pi t/T)/4$. In the following, we consider the contribution to the FI from the first four terms in \eqref{eq:FI_definition}  (contribution of modes with $n,m\leq 1$) as measurement distinguishing only a few most important modes is easier to implement and therefore it is of the highest interest.
Fig. \ref{fig:rot} investigates the FI for these two periodic motions. For both of these trajectories we see as expected based on (\ref{eq:FI_phi}) that for $\theta(t)=\pi/2$ FI  always lays between FI for the worst and best case scenario $\phi$ which both are free from Rayleigh's curse and thus converge to the optimal value $1/w^2$ in the limit $d\rightarrow 0$. Let us consider in more detail the FI for the case of rotations. One can calculate (see Appendix \ref{app:calc}) that the contribution to the FI from the first four terms in this case is given by: 
\begin{multline}
   w^2 F(d,\theta)=\frac{1}{8} \sin ^2(\theta ) e^{-x^2 \sin ^2(\theta )}\\ \left(x^6 \sin ^6(\theta )+4 x^4 \sin ^4(\theta )\right.
    \left.-4 x^2 \sin ^2(\theta )+8\right),
\end{multline}
where we introduced the notation $x:=d/2w$. As we already noted for $\theta(t)=\pi/2$, that is, when the system stays in the $xy$ plane, the FI goes with decreasing separation to the optimal value $1/w^2$. However, when $\theta(t)\neq \pi/2$ the FI decreases, since for a part of the measurement the effective separation of the sources in the imaging plane decreases. Therefore, the separation measurement becomes harder in this case because based on (\ref{eq:prec}) lower FI results in a worse scaling of the uncertainty with the number of registered photons $N$. Still, FI is nonzero for any $\theta(t)\neq 0, \pi$.

As a reference, we also consider the impact of the rotations on the emblematic measurement scheme, i.e., direct imaging. For the case where measurement is performed with perfect direct imaging, we have a qualitative difference in the FI if the brightnesses of the sources are unequal. If rotations are not present, then in such a case FI goes for small separations to a positive constant (but still suboptimal) dependent on $v$ \cite{Linowski}. However, when rotations are present, the behavior of FI changes to this associated with equal brightness case, as the FI for small separations decreases to zero independently from $v$ as $w^2 F_{D}\approx 4x^2\sin^4 \theta$   (see Appendix \ref{app:direct}). What is more, it performs even worse than in the stationary case with equal brightness of the sources ($v=1/2$) for which $w^2 F_{D}\approx 8x^2\sin^4 \theta$. Thus, in this scenario, the measurement based on SPADE gains an even greater advantage than in the stationary case.

\subsection{$\kappa=0,\,\phi(t)=const$}
Let us now analyze the scenario where $\phi(t)=const$ and $\theta(t)$ changes in time. This situation corresponds to a star transiting through the image of another star.  Fig. \ref{fig:rot} shows the FI in the case of rotation with constant angular velocity. For such a motion, FI for small separations has the approximate form:
\begin{equation}
  w^2 F(d,\phi) \approx\frac{1}{2}-\frac{1}{16} x^2 (3 \cos (4 \phi )+11).
\end{equation}
Rotation in the coordinate $\theta$ has a greater impact on FI as in the limit $x\rightarrow 0$ it approaches $1/2$. However, this is independent of $\phi$. The higher impact of such motion on FI results from the fact that the effective distance between the sources on the imaging plane changes in time, decreasing even to 0, and this varies probabilities much more than simple rotations in the $xy$ axis. Note that in the considered scenarios the angular velocity plays no particular role as long as the detection density is high enough or the time of measurement is much longer than the period of the rotation.

Similarly to the case of $\theta=const$ in this scenario, perfect direct imaging is also more prone to such dynamics as for small separations $w^2F_{D}(d)\approx2x^2$ with a modification by a factor of $1/4$ from the stationary case for $v=1/2$ and the qualitative difference for $v\neq1/2$ .
\subsection{Random orientation}
 In previous sections, we have checked the impact of the uniform random distribution separately of $\phi$ and $\theta$. Here, we investigate the scenario in which sources change their orientation randomly, resulting in a uniform distribution of orientations over the sphere: $\textbf{p}(\phi,\theta)=\sin(\theta)/4\pi$.
 
In Fig. \ref{fig:rot} the FI for this case is presented as a function of $x$. For the regarded distribution, FI has the following approximate form for small separations:
\begin{equation}
    w^2F(d)\approx\frac{2}{3}-\frac{8 x^2}{9}.
\end{equation}
Again, FI in the limit $x\rightarrow 0$ goes to a constant of the same order as the quantum Fisher information for the stationary problem for sources located parallel to the imaging plane, allowing efficient measurement of small separations. 

For this distribution, perfect direct imaging performs with the lowest precision as $w^2F_{D}(d)\approx16x^2/9$, which is, however, not the case for SPADE.
\subsection{$\kappa\neq0$}
Presented examples assumed $\kappa=0$ thus removing  any dependence on brightness $v$ and setting rotation axis in the middle between the sources. However, in astrophysical scenarios, the rotation axis is often shifted due to the difference in the masses of the objects. Let us now consider the system of two stars, with masses $M_1$ and $M_2$. In this case, the axis of rotation is shifted to the center of mass, and thus we can calculate:
\begin{equation}
    \kappa=\frac{M_2-M_1}{M_1+M_2}.
\end{equation}
However, this is not the only change, as the brightness of the star depends on its mass. Let us as an approximation calculate $v$ using the relation for the luminosity and mass for stars with mass $M\sim M_\odot$ where $M_\odot$ stands for the mass of the Sun. This relation reads \cite{duric2004advanced}:
\begin{equation}
    \frac{L_1}{L_2}=\left(\frac{M_1}{M_2}\right)^4=\frac{v}{1-v},
\end{equation}
what results in $v=M_1^4/(M_1^4+M_2^4)$. Here, we also use the fact that the distance from the apparatus is approximately the same for both sources, and thus intensity drop is by the same factor which results from inverse-square law. 

\begin{figure}
    \centering
    \includegraphics[width=\linewidth]{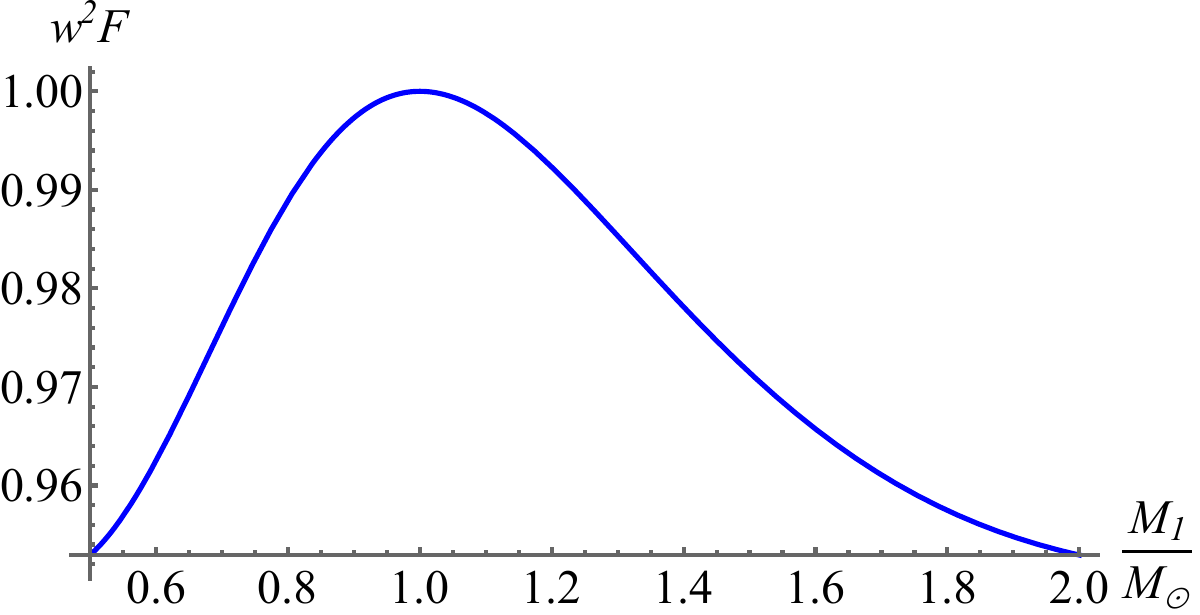}
    \caption{Small separation limit of Fisher information (\ref{eq:FI0}) for the system of two stars in relation to the mass of the first star $M_1$  assuming that $M_2=M_\odot$ and that the rotation is in the $xy$ plane. While the Fisher information is lowered for the imbalanced case $M_1\neq M_2$  it is still of the same order of magnitude as in the optimal case.}
    \label{fig:stars}
\end{figure}

Fig. \ref{fig:stars} presents a small separation limit of FI (\ref{eq:FI0}) in relation to the mass $M_1$ assuming $C=1$ and $M_2=M_\odot$. Note that this corresponds to the case $\theta(t)=\pi/2$ and arbitrary rotations in the $xy$ plane; however, for the case of rotations with constant angular velocity $\theta(t)=2\pi t/T$ we have $C=1/2$. We observe that FI decreases only slightly up to around $0.95$, and the maximum is obtained as expected for $M_1=M_2$. From this we see that the method is weakly affected by this change in the scenario and performs with near-optimal precision. 

An approximate FI can be calculated for direct imaging in the case where the rotation axis is shifted. These calculations do not yield any qualitative difference from the case of the centered rotation axis since for $\theta(t)=const,\, \phi(t)=2\pi t/T$ one gets $w^2F_D\approx 4x^2(1-\kappa+2\kappa v)\sin^4{\theta}$ and for the situation with the trajectories reversed $w^2F_D\approx 2x^2(1-\kappa+2\kappa v)$.
\section{Oscillations}
Consider, as another example, the case where the distance between sources oscillates in time. In such a case, we have to shift our attention from the estimation of the separation to the estimation of the average separation between sources $\overline{d}$, as the former is not well defined. For simplicity, let us set $\theta=\pi/2$, $\phi=0$ and $\kappa=0$. We will consider two cases, the first in which the amplitude of the oscillation is proportional to $\overline{d}$ with the trajectory given by $d(t)=\overline{d}(1+2 w A_1\cos{2\pi t/T})$ where $0<A_1<1$ is a parameter that we assume to know. The second case deals with arbitrary oscillations of set amplitude $A_2$, i.e., $d(t)=\overline{d}+2 w A_2\cos{2\pi t/T}$.
 
\begin{figure}[ht]
\centering
\includegraphics[width=0.49\textwidth]{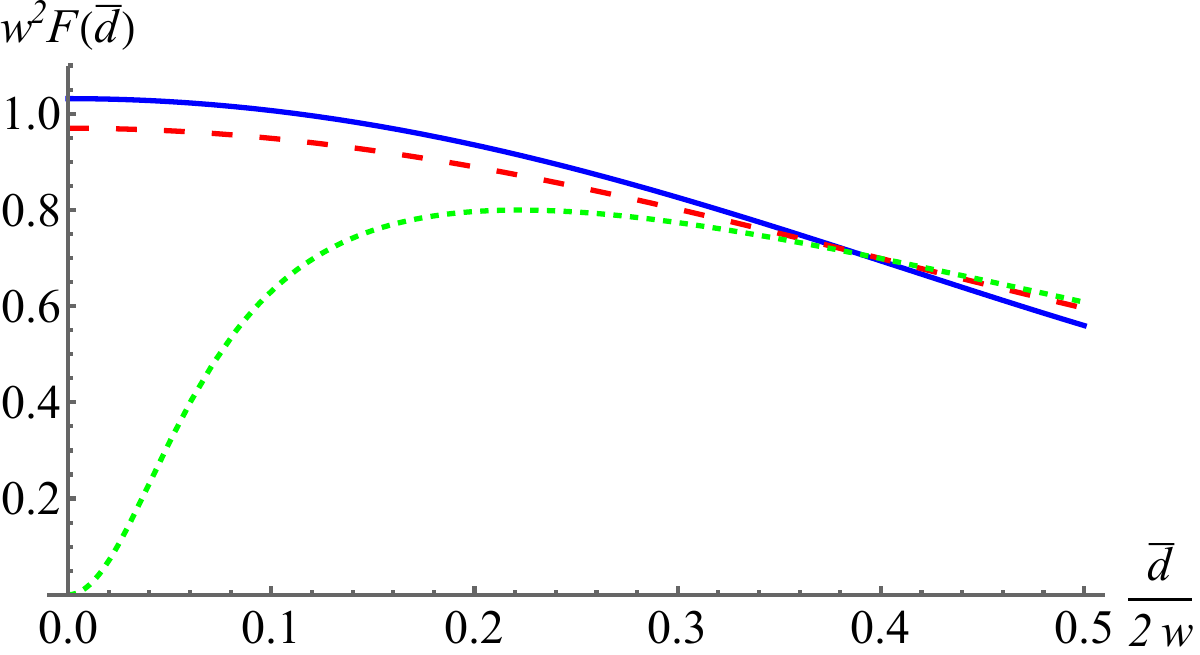} 
\caption{Fisher information for estimating average separation as a function of $\bar d/2w$ for different oscillation models. Oscillations with amplitude a priori dependent on average separation $\overline{d}$ where $d(t)=\overline{d}(1+2 w A_1\cos{2\pi t/T})$ with $A_1=1/4$ (blue curve), oscillations with fixed amplitude $d(t)=\overline{d}+2 w A_2\cos{2\pi t/T}$ for $A_2=0.1$ (green dashed curve) and $A_2=\overline{d}/4$ (red long dashed curve). Note that a priori knowledge of amplitude dependence allows for enhancement of precision of the scheme as for the such case FI exceeds $1$. For the fixed amplitude case, Fisher information is weakly affected by dynamics up to the point in which sources start to interchange their position, i.e. $A_2>\bar x$.  }
\label{fig:osc}
\end{figure}

 Fig. \ref{fig:osc} presents the FI for both oscillation models for $A_1=1/4$ and $A_2=0.1$ and also the amplitude that varies with a separation $A_2=\overline{d}A_1:=A_2'$. Note that the FI is calculated without specifying the amplitudes, and their values are put afterward. We can observe interestingly that assuming a priori dependence of oscillations on the average separation increases the precision and even surpasses the precision attainable for stationary sources. The approximate FI for such a case reads:
 \begin{equation}\label{eq:fi_osc1}
     w^2 F_{A1}(\bar d)\approx 1+\frac{A_1^2}{2}+\frac{1}{8} \left(-7 A_1^4-64
   A_1^2-16\right) \bar x^2,
 \end{equation}
where we put now $\bar x=\bar d/2w$. Note that higher amplitudes $A_1$ result in higher FI which increases to 2 when $A_1\rightarrow 1$. The reason for this is the fact that the change of $\overline{d}$ impacts the extra oscillatory degree of freedom, resulting in more pronounced differences in the probability distribution, and the scale of this impact is governed by $A_1$. This is opposite to the second model with $A_2'$ where the dependence on $\overline{d}$ is added afterward, as can be seen from the approximate expression for FI for $\bar x\ll 1$:
 \begin{equation}\label{eq:fi_osc2}
    w^2F_{A_2'}(\bar d)\approx \frac{2}{A_1^2+2}+\frac{\left(-19 A_1^4-32
   A_1^2-16\right) \bar x^2}{2 \left(A_1^2+2\right)^2}.
 \end{equation}
 For this model, increasing $A_1$ contributes only to the greater oscillations, which are not associated with $\overline{d}$ by any means in the model, thus adding ``noise'' for the estimation procedure. 
 In general, for the case of fixed amplitude of oscillations, FI has the following approximate forms:
 \begin{align} \label{eq:Fi_o}
    w^2 F_{A_2}(\bar d)\approx
    \begin{cases}
       1-2 \bar x^2+ A_2^2 \left(\bar x^2-\frac{1}{2 \bar x^2}-2\right)& \bar x\gg A_2, \\   
   \left(\frac{2}{A_2^2}-\frac{19}{2}+\frac{85 A_2^2}{8}\right) \bar x^2 & \bar x\ll A_2.
    \end{cases}
\end{align}
 We see that FI decreases from some point with decreasing $\bar x$ as the oscillations become comparable with $\bar x$. Note that for $\bar x<A_2$ the sources interchange their position and $\bar d$ has no longer interpretation as an average separation; thus if one is not interested in such a specific case, the drop in FI to $0$ with $\bar x\rightarrow0$ is not significant. 
 
In the case of regarded oscillations, the measurement using SPADE into Hermite-Gauss modes also outperforms perfect direct imaging as for direct imaging for small separations with equal brightness of the sources FI reads (see the Appendix
\ref{app:direct}): $w^2F_{A_1D}(\bar d)\approx2(2+A_1^2)^2\bar x^2$ and $w^2F_{A_2D}(\bar d)\approx8(1-4A_2^2)\bar x^2$.

 \section{Parameter reduction}
One of the problems with estimation tasks is that the precision often drops when more parameters need to be estimated, and additional parameters also add to the complexity of the estimation problem. When dealing with estimation of separation in sub-Rayleigh  regime, the orientation of the sources cannot in many cases be taken as a priori knowledge even if they are stationary. Although for extremely small separations, FI does not depend on the $\phi$ it can depend on it still in the sub-Rayleigh regime if one does not implement measurement in all Hermite-Gauss modes, and even when FI is not dependent on $\phi$ the probabilities are, and they are used in the estimation process.  Thus, in such cases, orientation may also require estimation. However, this problem can be partially solved by our results. The key feature that we will use is the fact that, for periodic motion, the dependence on initial condition disappears.

Having stationary or slowly moving sources, one can simply rotate an image with constant angular velocity throughout the measurement. This makes the stationary sources case reassemble the rotating sources case. Based on the results from section \ref{sec:theta} in this way one reduces the need to estimate the orientation angle $\phi$ without significant loss of precision in the comparison to the case where $\phi$ is a priori known. Thus, in general, if one is not interested in orientation, one obtains an easy algorithm, which simplifies the estimation and allows one to omit the precision losses due to multi-parameter estimation. One could also make a single slow rotation by $2\pi$ throughout the measurement, obtaining approximately the same result.

\section{Conclusions}
 In summary, we have presented how to treat dynamically changing systems of two weak incoherent sources in order to accurately perform parameter estimation with the photon-counting experiment in modes after spatial-mode demultiplexing. We have examined precision limits of separation estimation with spatial mode demultiplexing into Hermite-Gauss modes, for a relevant case from the astrophysical point of view of rotations, and as additional examples random orientation changes and oscillations. We have found that the estimation of the separation with SPADE is robust to dynamic changes of the orientation of the system and that it outperforms the estimation protocol, which utilizes perfect direct imaging. What is more, we have presented a method that allows one to reduce the need to estimate one of the orientation angles for stationary or slowly moving sources, thus obtaining a simpler and precise method of separation estimation. 
 
Our findings open the path for the implementation of spatial mode demultiplexing-based measurements in dynamical systems for parameter estimation. We provide a simple tool for analyzing the measurement scenario for the specific motion. Thus, if one is able to predict the type of motion for one's scenario, one can easily check if spatial mode demultiplexing provides an advantage over different methods. This allows for a more scenario-specific analysis of the problem in the future. Such an analysis could include different types of imperfections, like, e.g., displacement or general crosstalk. Note that different methods also need similar treatment to give accurate models for estimation, as we showed in the example of direct imaging. Thus, our findings can also impact research on different methods of parameter estimation than spatial mode demultiplexing.

One of the primary goals of passive imaging methods such as spatial mode demultiplexing considered here is the task of detecting exoplanets and binary stars. As we have found the particular robustness against rotations of separation measurement based on Hermite-Gauss modes, it would be interesting to see if this property is also transferred to the task of hypothesis testing to discriminate between one and two sources \cite{Hyp_1, Schlichtholz:24}.

 \section{ACKNOWLEDGMENTS}
Project ApresSF is supported by the National Science Centre (No. 2019/32/Z/ST2/00017), Poland, under QuantERA, which has received funding from the European Union's Horizon 2020 research and innovation programme under Grant Agreement No. 731473. 
\bibliography{biblio}

\begin{thebibliography}{36}%
\makeatletter
\providecommand \@ifxundefined [1]{%
 \@ifx{#1\undefined}
}%
\providecommand \@ifnum [1]{%
 \ifnum #1\expandafter \@firstoftwo
 \else \expandafter \@secondoftwo
 \fi
}%
\providecommand \@ifx [1]{%
 \ifx #1\expandafter \@firstoftwo
 \else \expandafter \@secondoftwo
 \fi
}%
\providecommand \natexlab [1]{#1}%
\providecommand \enquote  [1]{``#1''}%
\providecommand \bibnamefont  [1]{#1}%
\providecommand \bibfnamefont [1]{#1}%
\providecommand \citenamefont [1]{#1}%
\providecommand \href@noop [0]{\@secondoftwo}%
\providecommand \href [0]{\begingroup \@sanitize@url \@href}%
\providecommand \@href[1]{\@@startlink{#1}\@@href}%
\providecommand \@@href[1]{\endgroup#1\@@endlink}%
\providecommand \@sanitize@url [0]{\catcode `\\12\catcode `\$12\catcode
  `\&12\catcode `\#12\catcode `\^12\catcode `\_12\catcode `\%12\relax}%
\providecommand \@@startlink[1]{}%
\providecommand \@@endlink[0]{}%
\providecommand \url  [0]{\begingroup\@sanitize@url \@url }%
\providecommand \@url [1]{\endgroup\@href {#1}{\urlprefix }}%
\providecommand \urlprefix  [0]{URL }%
\providecommand \Eprint [0]{\href }%
\providecommand \doibase [0]{https://doi.org/}%
\providecommand \selectlanguage [0]{\@gobble}%
\providecommand \bibinfo  [0]{\@secondoftwo}%
\providecommand \bibfield  [0]{\@secondoftwo}%
\providecommand \translation [1]{[#1]}%
\providecommand \BibitemOpen [0]{}%
\providecommand \bibitemStop [0]{}%
\providecommand \bibitemNoStop [0]{.\EOS\space}%
\providecommand \EOS [0]{\spacefactor3000\relax}%
\providecommand \BibitemShut  [1]{\csname bibitem#1\endcsname}%
\let\auto@bib@innerbib\@empty
\bibitem [{\citenamefont {den Dekker}\ and\ \citenamefont {van~den
  Bos}(1997)}]{resolution_survey_denDekker_1997}%
  \BibitemOpen
  \bibfield  {author} {\bibinfo {author} {\bibfnamefont {A.~J.}\ \bibnamefont
  {den Dekker}}\ and\ \bibinfo {author} {\bibfnamefont {A.}~\bibnamefont
  {van~den Bos}},\ }\bibfield  {title} {\bibinfo {title} {Resolution: {A}
  survey},\ }\href {https://doi.org/10.1364/JOSAA.14.000547} {\bibfield
  {journal} {\bibinfo  {journal} {J. Opt. Soc. Am. A}\ }\textbf {\bibinfo
  {volume} {14}},\ \bibinfo {pages} {547} (\bibinfo {year} {1997})}\BibitemShut
  {NoStop}%
\bibitem [{\citenamefont {Goodman}(2005)}]{Fourier_optics_Goodman_2005}%
  \BibitemOpen
  \bibfield  {author} {\bibinfo {author} {\bibfnamefont {J.~W.}\ \bibnamefont
  {Goodman}},\ }\href@noop {} {\emph {\bibinfo {title} {Introduction to Fourier
  optics}}}\ (\bibinfo  {publisher} {Roberts and Company Publishers},\ \bibinfo
  {year} {2005})\BibitemShut {NoStop}%
\bibitem [{\citenamefont {Pa\'{u}r}\ \emph {et~al.}(2018)\citenamefont
  {Pa\'{u}r}, \citenamefont {Stoklasa}, \citenamefont {Grover}, \citenamefont
  {Krzic}, \citenamefont {S\'{a}nchez-Soto}, \citenamefont {Hradil},\ and\
  \citenamefont {\v{R}eh\'{a}\v{c}ek}}]{Rayleigh_curse_Paur_2018}%
  \BibitemOpen
  \bibfield  {author} {\bibinfo {author} {\bibfnamefont {M.}~\bibnamefont
  {Pa\'{u}r}}, \bibinfo {author} {\bibfnamefont {B.}~\bibnamefont {Stoklasa}},
  \bibinfo {author} {\bibfnamefont {J.}~\bibnamefont {Grover}}, \bibinfo
  {author} {\bibfnamefont {A.}~\bibnamefont {Krzic}}, \bibinfo {author}
  {\bibfnamefont {L.~L.}\ \bibnamefont {S\'{a}nchez-Soto}}, \bibinfo {author}
  {\bibfnamefont {Z.}~\bibnamefont {Hradil}},\ and\ \bibinfo {author}
  {\bibfnamefont {J.}~\bibnamefont {\v{R}eh\'{a}\v{c}ek}},\ }\bibfield  {title}
  {\bibinfo {title} {Tempering {Rayleigh's} curse with {PSF} shaping},\ }\href
  {https://doi.org/10.1364/OPTICA.5.001177} {\bibfield  {journal} {\bibinfo
  {journal} {Optica}\ }\textbf {\bibinfo {volume} {5}},\ \bibinfo {pages}
  {1177} (\bibinfo {year} {2018})}\BibitemShut {NoStop}%
\bibitem [{\citenamefont {Larson}\ and\ \citenamefont
  {Saleh}(2018)}]{Rayleigh_curse_Larson_2018}%
  \BibitemOpen
  \bibfield  {author} {\bibinfo {author} {\bibfnamefont {W.}~\bibnamefont
  {Larson}}\ and\ \bibinfo {author} {\bibfnamefont {B.~E.~A.}\ \bibnamefont
  {Saleh}},\ }\bibfield  {title} {\bibinfo {title} {Resurgence of {Rayleigh's}
  curse in the presence of partial coherence},\ }\href
  {https://doi.org/10.1364/OPTICA.5.001382} {\bibfield  {journal} {\bibinfo
  {journal} {Optica}\ }\textbf {\bibinfo {volume} {5}},\ \bibinfo {pages}
  {1382} (\bibinfo {year} {2018})}\BibitemShut {NoStop}%
\bibitem [{\citenamefont {Hell}(2007)}]{STORM_Hell_2007}%
  \BibitemOpen
  \bibfield  {author} {\bibinfo {author} {\bibfnamefont {S.~W.}\ \bibnamefont
  {Hell}},\ }\bibfield  {title} {\bibinfo {title} {Far-field optical
  nanoscopy},\ }\href {https://doi.org/10.1126/science.1137395} {\bibfield
  {journal} {\bibinfo  {journal} {Science}\ }\textbf {\bibinfo {volume}
  {316}},\ \bibinfo {pages} {1153} (\bibinfo {year} {2007})}\BibitemShut
  {NoStop}%
\bibitem [{\citenamefont {Smith}\ and\ \citenamefont
  {Gbur}(2016)}]{superoscillations_Smith_2016}%
  \BibitemOpen
  \bibfield  {author} {\bibinfo {author} {\bibfnamefont {M.~K.}\ \bibnamefont
  {Smith}}\ and\ \bibinfo {author} {\bibfnamefont {G.~J.}\ \bibnamefont
  {Gbur}},\ }\bibfield  {title} {\bibinfo {title} {Construction of arbitrary
  vortex and superoscillatory fields},\ }\href
  {https://doi.org/10.1364/OL.41.004979} {\bibfield  {journal} {\bibinfo
  {journal} {Opt. Lett.}\ }\textbf {\bibinfo {volume} {41}},\ \bibinfo {pages}
  {4979} (\bibinfo {year} {2016})}\BibitemShut {NoStop}%
\bibitem [{\citenamefont {Gbur}(2019)}]{superoscillations_Gbur_2019}%
  \BibitemOpen
  \bibfield  {author} {\bibinfo {author} {\bibfnamefont {G.}~\bibnamefont
  {Gbur}},\ }\bibfield  {title} {\bibinfo {title} {Using superoscillations for
  superresolved imaging and subwavelength focusing},\ }\href
  {https://doi.org/doi:10.1515/nanoph-2018-0112} {\bibfield  {journal}
  {\bibinfo  {journal} {Nanophotonics}\ }\textbf {\bibinfo {volume} {8}},\
  \bibinfo {pages} {205} (\bibinfo {year} {2019})}\BibitemShut {NoStop}%
\bibitem [{\citenamefont {Betzig}\ \emph {et~al.}(2006)\citenamefont {Betzig},
  \citenamefont {Patterson}, \citenamefont {Sougrat}, \citenamefont
  {Lindwasser}, \citenamefont {Olenych}, \citenamefont {Bonifacino},
  \citenamefont {Davidson}, \citenamefont {Lippincott-Schwartz},\ and\
  \citenamefont {Hess}}]{PALM_Betzig_2006}%
  \BibitemOpen
  \bibfield  {author} {\bibinfo {author} {\bibfnamefont {E.}~\bibnamefont
  {Betzig}}, \bibinfo {author} {\bibfnamefont {G.~H.}\ \bibnamefont
  {Patterson}}, \bibinfo {author} {\bibfnamefont {R.}~\bibnamefont {Sougrat}},
  \bibinfo {author} {\bibfnamefont {O.~W.}\ \bibnamefont {Lindwasser}},
  \bibinfo {author} {\bibfnamefont {S.}~\bibnamefont {Olenych}}, \bibinfo
  {author} {\bibfnamefont {J.~S.}\ \bibnamefont {Bonifacino}}, \bibinfo
  {author} {\bibfnamefont {M.~W.}\ \bibnamefont {Davidson}}, \bibinfo {author}
  {\bibfnamefont {J.}~\bibnamefont {Lippincott-Schwartz}},\ and\ \bibinfo
  {author} {\bibfnamefont {H.~F.}\ \bibnamefont {Hess}},\ }\bibfield  {title}
  {\bibinfo {title} {Imaging intracellular fluorescent proteins at nanometer
  resolution},\ }\href {https://doi.org/10.1126/science.1127344} {\bibfield
  {journal} {\bibinfo  {journal} {Science}\ }\textbf {\bibinfo {volume}
  {313}},\ \bibinfo {pages} {1642} (\bibinfo {year} {2006})}\BibitemShut
  {NoStop}%
\bibitem [{\citenamefont {Hess}\ \emph {et~al.}(2006)\citenamefont {Hess},
  \citenamefont {Girirajan},\ and\ \citenamefont {Mason}}]{PALM_Hess_2006}%
  \BibitemOpen
  \bibfield  {author} {\bibinfo {author} {\bibfnamefont {S.~T.}\ \bibnamefont
  {Hess}}, \bibinfo {author} {\bibfnamefont {T.~P.~K.}\ \bibnamefont
  {Girirajan}},\ and\ \bibinfo {author} {\bibfnamefont {M.~D.}\ \bibnamefont
  {Mason}},\ }\bibfield  {title} {\bibinfo {title} {Ultra-high resolution
  imaging by fluorescence photoactivation localization microscopy},\ }\href
  {https://doi.org/10.1529/biophysj.106.091116} {\bibfield  {journal} {\bibinfo
   {journal} {Biophys. J.}\ }\textbf {\bibinfo {volume} {91}},\ \bibinfo
  {pages} {4258} (\bibinfo {year} {2006})}\BibitemShut {NoStop}%
\bibitem [{\citenamefont {Hell}\ and\ \citenamefont
  {Wichmann}(1994)}]{Hell:94}%
  \BibitemOpen
  \bibfield  {author} {\bibinfo {author} {\bibfnamefont {S.~W.}\ \bibnamefont
  {Hell}}\ and\ \bibinfo {author} {\bibfnamefont {J.}~\bibnamefont
  {Wichmann}},\ }\bibfield  {title} {\bibinfo {title} {Breaking the diffraction
  resolution limit by stimulated emission: stimulated-emission-depletion
  fluorescence microscopy},\ }\href {https://doi.org/10.1364/OL.19.000780}
  {\bibfield  {journal} {\bibinfo  {journal} {Opt. Lett.}\ }\textbf {\bibinfo
  {volume} {19}},\ \bibinfo {pages} {780} (\bibinfo {year} {1994})}\BibitemShut
  {NoStop}%
\bibitem [{\citenamefont {Klar}\ \emph {et~al.}(2000)\citenamefont {Klar},
  \citenamefont {Jakobs}, \citenamefont {Dyba}, \citenamefont {Egner},\ and\
  \citenamefont {Hell}}]{Klar8206}%
  \BibitemOpen
  \bibfield  {author} {\bibinfo {author} {\bibfnamefont {T.~A.}\ \bibnamefont
  {Klar}}, \bibinfo {author} {\bibfnamefont {S.}~\bibnamefont {Jakobs}},
  \bibinfo {author} {\bibfnamefont {M.}~\bibnamefont {Dyba}}, \bibinfo {author}
  {\bibfnamefont {A.}~\bibnamefont {Egner}},\ and\ \bibinfo {author}
  {\bibfnamefont {S.~W.}\ \bibnamefont {Hell}},\ }\bibfield  {title} {\bibinfo
  {title} {Fluorescence microscopy with diffraction resolution barrier broken
  by stimulated emission},\ }\href {https://doi.org/10.1073/pnas.97.15.8206}
  {\bibfield  {journal} {\bibinfo  {journal} {Proceedings of the National
  Academy of Sciences}\ }\textbf {\bibinfo {volume} {97}},\ \bibinfo {pages}
  {8206} (\bibinfo {year} {2000})}\BibitemShut {NoStop}%
\bibitem [{\citenamefont {Tham}\ \emph {et~al.}(2017)\citenamefont {Tham},
  \citenamefont {Ferretti},\ and\ \citenamefont
  {Steinberg}}]{SPLICE_Tham_2017}%
  \BibitemOpen
  \bibfield  {author} {\bibinfo {author} {\bibfnamefont {W.-K.}\ \bibnamefont
  {Tham}}, \bibinfo {author} {\bibfnamefont {H.}~\bibnamefont {Ferretti}},\
  and\ \bibinfo {author} {\bibfnamefont {A.~M.}\ \bibnamefont {Steinberg}},\
  }\bibfield  {title} {\bibinfo {title} {Beating {Rayleigh's} curse by imaging
  using phase information},\ }\href
  {https://doi.org/10.1103/PhysRevLett.118.070801} {\bibfield  {journal}
  {\bibinfo  {journal} {Phys. Rev. Lett.}\ }\textbf {\bibinfo {volume} {118}},\
  \bibinfo {pages} {070801} (\bibinfo {year} {2017})}\BibitemShut {NoStop}%
\bibitem [{\citenamefont {Bonsma-Fisher}\ \emph {et~al.}(2019)\citenamefont
  {Bonsma-Fisher}, \citenamefont {Tham}, \citenamefont {Ferretti},\ and\
  \citenamefont {Steinberg}}]{SPLICE_Bonsma-Fisher_2019}%
  \BibitemOpen
  \bibfield  {author} {\bibinfo {author} {\bibfnamefont {K.~A.~G.}\
  \bibnamefont {Bonsma-Fisher}}, \bibinfo {author} {\bibfnamefont {W.-K.}\
  \bibnamefont {Tham}}, \bibinfo {author} {\bibfnamefont {H.}~\bibnamefont
  {Ferretti}},\ and\ \bibinfo {author} {\bibfnamefont {A.~M.}\ \bibnamefont
  {Steinberg}},\ }\bibfield  {title} {\bibinfo {title} {Realistic
  {sub-Rayleigh} imaging with phase-sensitive measurements},\ }\href
  {https://doi.org/10.1088/1367-2630/ab3d97} {\bibfield  {journal} {\bibinfo
  {journal} {New J. Phys.}\ }\textbf {\bibinfo {volume} {21}},\ \bibinfo
  {pages} {093010} (\bibinfo {year} {2019})}\BibitemShut {NoStop}%
\bibitem [{\citenamefont {Hsu}\ \emph {et~al.}(2004)\citenamefont {Hsu},
  \citenamefont {Delaubert}, \citenamefont {Lam},\ and\ \citenamefont
  {Bowen}}]{Hsu_2004}%
  \BibitemOpen
  \bibfield  {author} {\bibinfo {author} {\bibfnamefont {M.~T.~L.}\
  \bibnamefont {Hsu}}, \bibinfo {author} {\bibfnamefont {V.}~\bibnamefont
  {Delaubert}}, \bibinfo {author} {\bibfnamefont {P.~K.}\ \bibnamefont {Lam}},\
  and\ \bibinfo {author} {\bibfnamefont {W.~P.}\ \bibnamefont {Bowen}},\
  }\bibfield  {title} {\bibinfo {title} {Optimal optical measurement of small
  displacements},\ }\href {https://doi.org/10.1088/1464-4266/6/12/003}
  {\bibfield  {journal} {\bibinfo  {journal} {Journal of Optics B: Quantum and
  Semiclassical Optics}\ }\textbf {\bibinfo {volume} {6}},\ \bibinfo {pages}
  {495} (\bibinfo {year} {2004})}\BibitemShut {NoStop}%
\bibitem [{\citenamefont {Delaubert}\ \emph {et~al.}(2006)\citenamefont
  {Delaubert}, \citenamefont {Treps}, \citenamefont {Harb}, \citenamefont
  {Lam},\ and\ \citenamefont {Bachor}}]{Delaubert:06}%
  \BibitemOpen
  \bibfield  {author} {\bibinfo {author} {\bibfnamefont {V.}~\bibnamefont
  {Delaubert}}, \bibinfo {author} {\bibfnamefont {N.}~\bibnamefont {Treps}},
  \bibinfo {author} {\bibfnamefont {C.~C.}\ \bibnamefont {Harb}}, \bibinfo
  {author} {\bibfnamefont {P.~K.}\ \bibnamefont {Lam}},\ and\ \bibinfo {author}
  {\bibfnamefont {H.-A.}\ \bibnamefont {Bachor}},\ }\bibfield  {title}
  {\bibinfo {title} {Quantum measurements of spatial conjugate variables:
  displacement and tilt of a gaussian beam},\ }\href
  {https://doi.org/10.1364/OL.31.001537} {\bibfield  {journal} {\bibinfo
  {journal} {Opt. Lett.}\ }\textbf {\bibinfo {volume} {31}},\ \bibinfo {pages}
  {1537} (\bibinfo {year} {2006})}\BibitemShut {NoStop}%
\bibitem [{\citenamefont {Tsang}\ \emph {et~al.}(2016)\citenamefont {Tsang},
  \citenamefont {Nair},\ and\ \citenamefont {Lu}}]{superresolution_Tsang_2016}%
  \BibitemOpen
  \bibfield  {author} {\bibinfo {author} {\bibfnamefont {M.}~\bibnamefont
  {Tsang}}, \bibinfo {author} {\bibfnamefont {R.}~\bibnamefont {Nair}},\ and\
  \bibinfo {author} {\bibfnamefont {X.-M.}\ \bibnamefont {Lu}},\ }\bibfield
  {title} {\bibinfo {title} {Quantum theory of superresolution for two
  incoherent optical point sources},\ }\href
  {https://doi.org/10.1103/PhysRevX.6.031033} {\bibfield  {journal} {\bibinfo
  {journal} {Phys. Rev. X}\ }\textbf {\bibinfo {volume} {6}},\ \bibinfo {pages}
  {031033} (\bibinfo {year} {2016})}\BibitemShut {NoStop}%
\bibitem [{\citenamefont {Tsang}(2019)}]{superresolution_starlight_Tsang_2019}%
  \BibitemOpen
  \bibfield  {author} {\bibinfo {author} {\bibfnamefont {M.}~\bibnamefont
  {Tsang}},\ }\bibfield  {title} {\bibinfo {title} {Resolving starlight: a
  quantum perspective},\ }\href {https://doi.org/10.1080/00107514.2020.1736375}
  {\bibfield  {journal} {\bibinfo  {journal} {Contemp. Phys.}\ }\textbf
  {\bibinfo {volume} {60}},\ \bibinfo {pages} {279} (\bibinfo {year}
  {2019})}\BibitemShut {NoStop}%
\bibitem [{\citenamefont {Santamaria}\ \emph {et~al.}(2023)\citenamefont
  {Santamaria}, \citenamefont {Pallotti}, \citenamefont {de~Cumis},
  \citenamefont {Dequal},\ and\ \citenamefont {Lupo}}]{Santamaria:23}%
  \BibitemOpen
  \bibfield  {author} {\bibinfo {author} {\bibfnamefont {L.}~\bibnamefont
  {Santamaria}}, \bibinfo {author} {\bibfnamefont {D.}~\bibnamefont
  {Pallotti}}, \bibinfo {author} {\bibfnamefont {M.~S.}\ \bibnamefont
  {de~Cumis}}, \bibinfo {author} {\bibfnamefont {D.}~\bibnamefont {Dequal}},\
  and\ \bibinfo {author} {\bibfnamefont {C.}~\bibnamefont {Lupo}},\ }\bibfield
  {title} {\bibinfo {title} {Spatial-mode demultiplexing for enhanced intensity
  and distance measurement},\ }\href {https://doi.org/10.1364/OE.486617}
  {\bibfield  {journal} {\bibinfo  {journal} {Opt. Express}\ }\textbf {\bibinfo
  {volume} {31}},\ \bibinfo {pages} {33930} (\bibinfo {year}
  {2023})}\BibitemShut {NoStop}%
\bibitem [{\citenamefont {Santamaria}\ \emph {et~al.}(2024)\citenamefont
  {Santamaria}, \citenamefont {Sgobba},\ and\ \citenamefont
  {Lupo}}]{Santamaria:24}%
  \BibitemOpen
  \bibfield  {author} {\bibinfo {author} {\bibfnamefont {L.}~\bibnamefont
  {Santamaria}}, \bibinfo {author} {\bibfnamefont {F.}~\bibnamefont {Sgobba}},\
  and\ \bibinfo {author} {\bibfnamefont {C.}~\bibnamefont {Lupo}},\ }\bibfield
  {title} {\bibinfo {title} {Single-photon sub-rayleigh precision measurements
  of a pair of incoherent sources of unequal intensity},\ }\href
  {https://doi.org/10.1364/OPTICAQ.505457} {\bibfield  {journal} {\bibinfo
  {journal} {Optica Quantum}\ }\textbf {\bibinfo {volume} {2}},\ \bibinfo
  {pages} {46} (\bibinfo {year} {2024})}\BibitemShut {NoStop}%
\bibitem [{\citenamefont {Rouvi\`{e}re}\ \emph {et~al.}(2024)\citenamefont
  {Rouvi\`{e}re}, \citenamefont {Barral}, \citenamefont {Grateau},
  \citenamefont {Karuseichyk}, \citenamefont {Sorelli}, \citenamefont
  {Walschaers},\ and\ \citenamefont {Treps}}]{Rouviere:24}%
  \BibitemOpen
  \bibfield  {author} {\bibinfo {author} {\bibfnamefont {C.}~\bibnamefont
  {Rouvi\`{e}re}}, \bibinfo {author} {\bibfnamefont {D.}~\bibnamefont
  {Barral}}, \bibinfo {author} {\bibfnamefont {A.}~\bibnamefont {Grateau}},
  \bibinfo {author} {\bibfnamefont {I.}~\bibnamefont {Karuseichyk}}, \bibinfo
  {author} {\bibfnamefont {G.}~\bibnamefont {Sorelli}}, \bibinfo {author}
  {\bibfnamefont {M.}~\bibnamefont {Walschaers}},\ and\ \bibinfo {author}
  {\bibfnamefont {N.}~\bibnamefont {Treps}},\ }\bibfield  {title} {\bibinfo
  {title} {Ultra-sensitive separation estimation of optical sources},\ }\href
  {https://doi.org/10.1364/OPTICA.500039} {\bibfield  {journal} {\bibinfo
  {journal} {Optica}\ }\textbf {\bibinfo {volume} {11}},\ \bibinfo {pages}
  {166} (\bibinfo {year} {2024})}\BibitemShut {NoStop}%
\bibitem [{\citenamefont {Len}\ \emph {et~al.}(2020)\citenamefont {Len},
  \citenamefont {Datta}, \citenamefont {Parniak},\ and\ \citenamefont
  {Banaszek}}]{Len}%
  \BibitemOpen
  \bibfield  {author} {\bibinfo {author} {\bibfnamefont {Y.~L.}\ \bibnamefont
  {Len}}, \bibinfo {author} {\bibfnamefont {C.}~\bibnamefont {Datta}}, \bibinfo
  {author} {\bibfnamefont {M.}~\bibnamefont {Parniak}},\ and\ \bibinfo {author}
  {\bibfnamefont {K.}~\bibnamefont {Banaszek}},\ }\bibfield  {title} {\bibinfo
  {title} {Resolution limits of spatial mode demultiplexing with noisy
  detection},\ }\href {https://doi.org/10.1142/S0219749919410156} {\bibfield
  {journal} {\bibinfo  {journal} {International Journal of Quantum
  Information}\ }\textbf {\bibinfo {volume} {18}},\ \bibinfo {pages} {1941015}
  (\bibinfo {year} {2020})}\BibitemShut {NoStop}%
\bibitem [{\citenamefont {Ko{\l}ody{\'{n}}ski}\ and\ \citenamefont
  {Demkowicz-Dobrza{\'{n}}ski}(2013)}]{metrology_noise_Kolodynski_2013}%
  \BibitemOpen
  \bibfield  {author} {\bibinfo {author} {\bibfnamefont {J.}~\bibnamefont
  {Ko{\l}ody{\'{n}}ski}}\ and\ \bibinfo {author} {\bibfnamefont
  {R.}~\bibnamefont {Demkowicz-Dobrza{\'{n}}ski}},\ }\bibfield  {title}
  {\bibinfo {title} {Efficient tools for quantum metrology with uncorrelated
  noise},\ }\href {https://doi.org/10.1088/1367-2630/15/7/073043} {\bibfield
  {journal} {\bibinfo  {journal} {New J. Phys.}\ }\textbf {\bibinfo {volume}
  {15}},\ \bibinfo {pages} {073043} (\bibinfo {year} {2013})}\BibitemShut
  {NoStop}%
\bibitem [{\citenamefont {Gessner}\ \emph {et~al.}(2020)\citenamefont
  {Gessner}, \citenamefont {Fabre},\ and\ \citenamefont {Treps}}]{Manuel}%
  \BibitemOpen
  \bibfield  {author} {\bibinfo {author} {\bibfnamefont {M.}~\bibnamefont
  {Gessner}}, \bibinfo {author} {\bibfnamefont {C.}~\bibnamefont {Fabre}},\
  and\ \bibinfo {author} {\bibfnamefont {N.}~\bibnamefont {Treps}},\ }\bibfield
   {title} {\bibinfo {title} {Superresolution limits from measurement
  crosstalk},\ }\href {https://doi.org/10.1103/PhysRevLett.125.100501}
  {\bibfield  {journal} {\bibinfo  {journal} {Phys. Rev. Lett.}\ }\textbf
  {\bibinfo {volume} {125}},\ \bibinfo {pages} {100501} (\bibinfo {year}
  {2020})}\BibitemShut {NoStop}%
\bibitem [{\citenamefont {Sorelli}\ \emph {et~al.}(2021)\citenamefont
  {Sorelli}, \citenamefont {Gessner}, \citenamefont {Walschaers},\ and\
  \citenamefont {Treps}}]{Giacomo}%
  \BibitemOpen
  \bibfield  {author} {\bibinfo {author} {\bibfnamefont {G.}~\bibnamefont
  {Sorelli}}, \bibinfo {author} {\bibfnamefont {M.}~\bibnamefont {Gessner}},
  \bibinfo {author} {\bibfnamefont {M.}~\bibnamefont {Walschaers}},\ and\
  \bibinfo {author} {\bibfnamefont {N.}~\bibnamefont {Treps}},\ }\bibfield
  {title} {\bibinfo {title} {Moment-based superresolution: Formalism and
  applications},\ }\href {https://doi.org/10.1103/PhysRevA.104.033515}
  {\bibfield  {journal} {\bibinfo  {journal} {Phys. Rev. A}\ }\textbf {\bibinfo
  {volume} {104}},\ \bibinfo {pages} {033515} (\bibinfo {year}
  {2021})}\BibitemShut {NoStop}%
\bibitem [{\citenamefont {Linowski}\ \emph {et~al.}(2023)\citenamefont
  {Linowski}, \citenamefont {Schlichtholz}, \citenamefont {Sorelli},
  \citenamefont {Gessner}, \citenamefont {Walschaers}, \citenamefont {Treps},\
  and\ \citenamefont {Rudnicki}}]{Linowski}%
  \BibitemOpen
  \bibfield  {author} {\bibinfo {author} {\bibfnamefont {T.}~\bibnamefont
  {Linowski}}, \bibinfo {author} {\bibfnamefont {K.}~\bibnamefont
  {Schlichtholz}}, \bibinfo {author} {\bibfnamefont {G.}~\bibnamefont
  {Sorelli}}, \bibinfo {author} {\bibfnamefont {M.}~\bibnamefont {Gessner}},
  \bibinfo {author} {\bibfnamefont {M.}~\bibnamefont {Walschaers}}, \bibinfo
  {author} {\bibfnamefont {N.}~\bibnamefont {Treps}},\ and\ \bibinfo {author}
  {\bibfnamefont {{\L}.}~\bibnamefont {Rudnicki}},\ }\bibfield  {title}
  {\bibinfo {title} {Application range of crosstalk-affected spatial
  demultiplexing for resolving separations between unbalanced sources},\ }\href
  {https://doi.org/10.1088/1367-2630/ad0173} {\bibfield  {journal} {\bibinfo
  {journal} {New J. Phys.}\ }\textbf {\bibinfo {volume} {25}},\ \bibinfo
  {pages} {103050} (\bibinfo {year} {2023})}\BibitemShut {NoStop}%
\bibitem [{\citenamefont {Huang}\ and\ \citenamefont
  {Lupo}(2021)}]{hypothesis_testing_exoplanets_2021}%
  \BibitemOpen
  \bibfield  {author} {\bibinfo {author} {\bibfnamefont {Z.}~\bibnamefont
  {Huang}}\ and\ \bibinfo {author} {\bibfnamefont {C.}~\bibnamefont {Lupo}},\
  }\bibfield  {title} {\bibinfo {title} {Quantum hypothesis testing for
  exoplanet detection},\ }\href
  {https://doi.org/10.1103/PhysRevLett.127.130502} {\bibfield  {journal}
  {\bibinfo  {journal} {Phys. Rev. Lett.}\ }\textbf {\bibinfo {volume} {127}},\
  \bibinfo {pages} {130502} (\bibinfo {year} {2021})}\BibitemShut {NoStop}%
\bibitem [{\citenamefont {Zanforlin}\ \emph {et~al.}(2022)\citenamefont
  {Zanforlin}, \citenamefont {Lupo}, \citenamefont {Connolly}, \citenamefont
  {Kok}, \citenamefont {Buller},\ and\ \citenamefont {Huang}}]{Exo_5}%
  \BibitemOpen
  \bibfield  {author} {\bibinfo {author} {\bibfnamefont {U.}~\bibnamefont
  {Zanforlin}}, \bibinfo {author} {\bibfnamefont {C.}~\bibnamefont {Lupo}},
  \bibinfo {author} {\bibfnamefont {P.~W.~R.}\ \bibnamefont {Connolly}},
  \bibinfo {author} {\bibfnamefont {P.}~\bibnamefont {Kok}}, \bibinfo {author}
  {\bibfnamefont {G.~S.}\ \bibnamefont {Buller}},\ and\ \bibinfo {author}
  {\bibfnamefont {Z.}~\bibnamefont {Huang}},\ }\bibfield  {title} {\bibinfo
  {title} {Optical quantum super-resolution imaging and hypothesis testing},\
  }\href {https://doi.org/10.1038/s41467-022-32977-8} {\bibfield  {journal}
  {\bibinfo  {journal} {Nature Communications}\ }\textbf {\bibinfo {volume}
  {13}},\ \bibinfo {pages} {5373} (\bibinfo {year} {2022})}\BibitemShut
  {NoStop}%
\bibitem [{\citenamefont {Lu}\ \emph {et~al.}(2018)\citenamefont {Lu},
  \citenamefont {Krovi}, \citenamefont {Nair}, \citenamefont {Guha},\ and\
  \citenamefont {Shapiro}}]{Hyp_1}%
  \BibitemOpen
  \bibfield  {author} {\bibinfo {author} {\bibfnamefont {X.-M.}\ \bibnamefont
  {Lu}}, \bibinfo {author} {\bibfnamefont {H.}~\bibnamefont {Krovi}}, \bibinfo
  {author} {\bibfnamefont {R.}~\bibnamefont {Nair}}, \bibinfo {author}
  {\bibfnamefont {S.}~\bibnamefont {Guha}},\ and\ \bibinfo {author}
  {\bibfnamefont {J.~H.}\ \bibnamefont {Shapiro}},\ }\bibfield  {title}
  {\bibinfo {title} {Quantum-optimal detection of one-versus-two incoherent
  optical sources with arbitrary separation},\ }\href
  {https://doi.org/10.1038/s41534-018-0114-y} {\bibfield  {journal} {\bibinfo
  {journal} {npj Quantum Information}\ }\textbf {\bibinfo {volume} {4}},\
  \bibinfo {pages} {64} (\bibinfo {year} {2018})}\BibitemShut {NoStop}%
\bibitem [{\citenamefont {Schlichtholz}\ \emph {et~al.}(2024)\citenamefont
  {Schlichtholz}, \citenamefont {Linowski}, \citenamefont {Walschaers},
  \citenamefont {Treps}, \citenamefont {Rudnicki},\ and\ \citenamefont
  {Sorelli}}]{Schlichtholz:24}%
  \BibitemOpen
  \bibfield  {author} {\bibinfo {author} {\bibfnamefont {K.}~\bibnamefont
  {Schlichtholz}}, \bibinfo {author} {\bibfnamefont {T.}~\bibnamefont
  {Linowski}}, \bibinfo {author} {\bibfnamefont {M.}~\bibnamefont
  {Walschaers}}, \bibinfo {author} {\bibfnamefont {N.}~\bibnamefont {Treps}},
  \bibinfo {author} {\bibfnamefont {{\L}.}~\bibnamefont {Rudnicki}},\ and\
  \bibinfo {author} {\bibfnamefont {G.}~\bibnamefont {Sorelli}},\ }\bibfield
  {title} {\bibinfo {title} {Practical tests for sub-{R}ayleigh source
  discriminations with imperfect demultiplexers},\ }\href
  {https://doi.org/10.1364/OPTICAQ.502459} {\bibfield  {journal} {\bibinfo
  {journal} {Optica Quantum}\ }\textbf {\bibinfo {volume} {2}},\ \bibinfo
  {pages} {29} (\bibinfo {year} {2024})}\BibitemShut {NoStop}%
\bibitem [{\citenamefont {{Fischer}}\ \emph {et~al.}(2014)\citenamefont
  {{Fischer}}, \citenamefont {{Howard}}, \citenamefont {{Laughlin}},
  \citenamefont {{Macintosh}}, \citenamefont {{Mahadevan}}, \citenamefont
  {{Sahlmann}},\ and\ \citenamefont {{Yee}}}]{Exo_te}%
  \BibitemOpen
  \bibfield  {author} {\bibinfo {author} {\bibfnamefont {D.~A.}\ \bibnamefont
  {{Fischer}}}, \bibinfo {author} {\bibfnamefont {A.~W.}\ \bibnamefont
  {{Howard}}}, \bibinfo {author} {\bibfnamefont {G.~P.}\ \bibnamefont
  {{Laughlin}}}, \bibinfo {author} {\bibfnamefont {B.}~\bibnamefont
  {{Macintosh}}}, \bibinfo {author} {\bibfnamefont {S.}~\bibnamefont
  {{Mahadevan}}}, \bibinfo {author} {\bibfnamefont {J.}~\bibnamefont
  {{Sahlmann}}},\ and\ \bibinfo {author} {\bibfnamefont {J.~C.}\ \bibnamefont
  {{Yee}}},\ }\bibfield  {title} {\bibinfo {title} {{Exoplanet Detection
  Techniques}},\ }in\ \href
  {https://doi.org/10.2458/azu_uapress_9780816531240-ch031} {\emph {\bibinfo
  {booktitle} {Protostars and Planets VI}}},\ \bibinfo {editor} {edited by\
  \bibinfo {editor} {\bibfnamefont {H.}~\bibnamefont {{Beuther}}}, \bibinfo
  {editor} {\bibfnamefont {R.~S.}\ \bibnamefont {{Klessen}}}, \bibinfo {editor}
  {\bibfnamefont {C.~P.}\ \bibnamefont {{Dullemond}}},\ and\ \bibinfo {editor}
  {\bibfnamefont {T.}~\bibnamefont {{Henning}}}}\ (\bibinfo {year} {2014})\
  pp.\ \bibinfo {pages} {715--737},\ \Eprint {https://arxiv.org/abs/1505.06869}
  {arXiv:1505.06869 [astro-ph.EP]} \BibitemShut {NoStop}%
\bibitem [{\citenamefont {Goodman}(1985)}]{goodman1985}%
  \BibitemOpen
  \bibfield  {author} {\bibinfo {author} {\bibfnamefont {J.~W.}\ \bibnamefont
  {Goodman}},\ }\href@noop {} {\emph {\bibinfo {title} {Statistical optics}}}\
  (\bibinfo  {publisher} {Wiley},\ \bibinfo {address} {New York},\ \bibinfo
  {year} {1985})\BibitemShut {NoStop}%
\bibitem [{\citenamefont {Kay}(1993)}]{Kay}%
  \BibitemOpen
  \bibfield  {author} {\bibinfo {author} {\bibfnamefont {S.~M.}\ \bibnamefont
  {Kay}},\ }\href@noop {} {\emph {\bibinfo {title} {Fundamentals of Statistical
  Signal Processing}}}\ (\bibinfo  {publisher} {Prentice-Hall PTR, Upper Saddle
  River, NJ},\ \bibinfo {address} {New York},\ \bibinfo {year}
  {1993})\BibitemShut {NoStop}%
\bibitem [{\citenamefont {Chao}\ \emph {et~al.}(2016)\citenamefont {Chao},
  \citenamefont {Ward},\ and\ \citenamefont {Ober}}]{Chao:16}%
  \BibitemOpen
  \bibfield  {author} {\bibinfo {author} {\bibfnamefont {J.}~\bibnamefont
  {Chao}}, \bibinfo {author} {\bibfnamefont {E.~S.}\ \bibnamefont {Ward}},\
  and\ \bibinfo {author} {\bibfnamefont {R.~J.}\ \bibnamefont {Ober}},\
  }\bibfield  {title} {\bibinfo {title} {Fisher information theory for
  parameter estimation in single molecule microscopy: tutorial},\ }\href
  {https://doi.org/10.1364/JOSAA.33.000B36} {\bibfield  {journal} {\bibinfo
  {journal} {J. Opt. Soc. Am. A}\ }\textbf {\bibinfo {volume} {33}},\ \bibinfo
  {pages} {B36} (\bibinfo {year} {2016})}\BibitemShut {NoStop}%
\bibitem [{\citenamefont {Ruelle}(1979)}]{Ruelle1979}%
  \BibitemOpen
  \bibfield  {author} {\bibinfo {author} {\bibfnamefont {D.}~\bibnamefont
  {Ruelle}},\ }\bibfield  {title} {\bibinfo {title} {Ergodic theory of
  differentiable dynamical systems},\ }\href
  {https://doi.org/10.1007/BF02684768} {\bibfield  {journal} {\bibinfo
  {journal} {Publications Math{\'e}matiques de l'Institut des Hautes {\'E}tudes
  Scientifiques}\ }\textbf {\bibinfo {volume} {50}},\ \bibinfo {pages} {27}
  (\bibinfo {year} {1979})}\BibitemShut {NoStop}%
\bibitem [{\citenamefont {Moore}(2015)}]{Moore_Ergodic}%
  \BibitemOpen
  \bibfield  {author} {\bibinfo {author} {\bibfnamefont {C.~C.}\ \bibnamefont
  {Moore}},\ }\bibfield  {title} {\bibinfo {title} {Ergodic theorem, ergodic
  theory, and statistical mechanics},\ }\href
  {https://doi.org/10.1073/pnas.1421798112} {\bibfield  {journal} {\bibinfo
  {journal} {Proceedings of the National Academy of Sciences}\ }\textbf
  {\bibinfo {volume} {112}},\ \bibinfo {pages} {1907} (\bibinfo {year}
  {2015})}\BibitemShut {NoStop}%
\bibitem [{\citenamefont {Duric}(2004)}]{duric2004advanced}%
  \BibitemOpen
  \bibfield  {author} {\bibinfo {author} {\bibfnamefont {N.}~\bibnamefont
  {Duric}},\ }\href@noop {} {\emph {\bibinfo {title} {Advanced astrophysics}}}\
  (\bibinfo  {publisher} {Cambridge University Press},\ \bibinfo {year}
  {2004})\BibitemShut {NoStop}%
\end{thebibliography}%
 
\appendix
\section{Fisher information calculations}\label{app:calc}
\setcounter{equation}{0}

We recall from \cite{Linowski} that for the system of two weak incoherent sources separated by distance $d$ the probability of measuring the photon in $(n,m)$ Hermite-Gauss mode is given by:
\begin{equation}
    p(nm|\vec{r}_i,v,d)=v| f_{nm}(\vec{r}_1)|^2+(1-v)| f_{nm}(\vec{r}_2)|^2,\label{eq:prob_gen}
\end{equation}
where $\vec{r}_i=(r_i \cos\phi_i,r_i\sin\phi_i)$ denotes the position of the $i-th$ source in the image plane and
\begin{equation}
 f_{nm}(\vec{r}_i)=\int_{\mathbf{R}^2} d\vec{r}u_{nm}^*(r)u_{00}(\vec{r}-\vec{r}_i).
\end{equation}
are overlap integrals between Hermite-Gauss modes and spatial distribution of field from $i-th$ source. Hermite-Gauss basis functions can be put as:
\begin{multline}
u_{nm}(\vec{r})= \frac{\exp{-((r\cos\phi)^2+(r\sin\phi)^2)/w^2}}{\sqrt{(\pi/2)w^2 2^{n+m}n!m!}}\\ \times H_n\left(\sqrt{2}\frac{r\cos\phi}{w}\right)  H_m\left(\sqrt{2}\frac{r\sin\phi}{w}\right),  
\end{multline} 
where $H_n$ denotes Hermite polynomials and $w$ is the width of the point spread function. If Hermite-Gauss modes are centered in the middle between the sources, we have $\vec{r}_2=-\vec{r}_1$ and those overlap integrals read:
\begin{multline} \label{eq:beta_nm}
    \beta_{\pm nm} =f_{nm}(\pm\vec{r}_1) \\=\frac{1}{\sqrt{n!m!}}\left(\pm \frac{r_1}{w}\right)^{n+m} 
        \cos^n\phi \, \sin^m\phi \, e^{-r_1^2/2w^2},
\end{multline}
where we have dropped the subscript in $\phi_1$. From this and from Eq.  (\ref{eq:prob_gen}) follows that the probability of measuring a photon in the $(n,m)$ mode reads: 
\begin{equation}
    p(nm|\vec r_i,v,d)=\beta_{+ nm}^2.\label{eq:pr_nm}
\end{equation}

For rotations in three dimensions (that is, when the axis going through the sources is not parallel to the imaging plane), one has $r_{1(2)}=\frac{d}{2}\sin\theta $, where $d$ stands for the separation between sources and $\theta$ is the second orientation angle. In addition, to shift the origin out of the center of the system by $\xi$ but leave it between the sources, one modifies $r_{1(2)}$  by $d\rightarrow d\pm 2\xi w$ with plus for the first source and minus for the second. Thus, denoting probability for this general case as $p(nm|d,\phi,\theta,v,\xi)$, one obtains:
\begin{equation}\label{app:pnm-rot}
    p(nm|d,v,\xi)=\int_0^{2\pi} d\phi\int_{0}^{\pi}d\theta\, \textbf{p}(\phi,\theta) p(nm|d,\phi,\theta,v,\xi),
\end{equation}
and for a periodic trajectory, one can alternatively write:
\begin{equation}
    p(nm|d,v,\xi)=\frac{1}{T}\int_0^{T} dt\,p(nm|d,\phi(t),\theta(t),v,\xi).
\end{equation}

For oscillations,  we consider $\xi=0$ and $r_{1(2)}=(\overline{d}+a)\sin(\theta)/2 $, where $a$ describes the displacement from the average separation. From this follows
\begin{multline}
    p(nm|\overline{d},\phi,\theta)=\int_{- \infty}^{\infty} d a\,\textbf{p}(a) p(nm|\overline{d}+a,\phi,\theta,v,0)\\=\frac{1}{T}\int_{0}^{T} dt\,p(nm|\overline{d}+a(t),\phi,\theta,v,0).
\end{multline}

Inserting those probabilities into (\ref{eq:FI_definition}) one obtains Fisher information for specific scenarios. The approximate forms of FI $d\ll1$ for the case of rotations if $\xi=0$ are obtained by expanding FI into the power series in $d$, or by first expanding $p(nm|d,\phi,\theta,v,\xi)$ if the integrals have no known analytical form and then expanding FI. If $\xi\neq0$ one makes a modification before expansion $d\rightarrow\epsilon d$ and $\xi\rightarrow\epsilon\xi$, and expands in powers of $\epsilon$ setting it afterward to 1.  This is done to account for the fact that both these parameters can be comparable. The case of oscillations calculation of approximate FI  (\ref {eq:fi_osc1}) and (\ref {eq:fi_osc1}) is analogous, however, we first take the leading orders of the series expansion of $p(nm|d=\bar d+a,\phi,\theta,v,0)$ in terms of $d$ and then perform the series expansion of the resultant FI in the leading orders of $\bar d$.  To calculate FI  (\ref{eq:Fi_o}) for the scenario with set oscillation amplitude $A_2$ for the case of $x\gg A_2$ we expand FI into series first in $\bar d$ and then in $A_2$ and for the opposite case $x\ll A_2$ we reverse the order of expansions.

\section{Proof of the Proposition 1: Robustness against rotations}\label{app:t0}
Let us analyze the impact of dynamic orientation angles on FI for estimating separation using the measurement in the Hermite-Gauss basis. In the scenario considered, one can write the term of FI coming from the mode (n,m) in terms of probability (\ref{app:pnm-rot}) as:
\begin{multline}
w^2F_{nm}(d,v,\xi)=\frac{(\frac{d}{dx} p(nm|x,v,\xi))^2}{4 p(nm|x,v,\xi)}\\=\frac{\Big(\frac{d}{dx} \int_0^{2\pi}d\phi \int_{0}^{\pi}d\theta\, \textbf{p}(\phi,\theta)p(n,m|x,\phi,\theta,v,\xi)\Big)^2}{4 p(nm
|x,v,\xi)},
\end{multline}
here we once again use $x=d/2w$. If the probability distribution $\textbf{p}(\phi,\theta)$ behaves well enough, we can interchange the order of the derivative with the integrals.  The important realization is that for $x\ll 1$ probability $p(n,m|x,\phi,\theta,v,\xi)$ approximately factorizes into a part dependent on angles $c_{nm}(\phi,\theta)$ and a remaining part $f_{nm}(x,v,\xi)$. This is because the leading orders of expansion in $x$ of this probability (expansion is obtained through the small parameter $\epsilon$ as described in the previous section and then setting $\xi=\kappa x$) have the form :
\begin{multline}
    p(n,m|x,\phi,\theta,v,\xi)\approx \big[\cos^{2m}(\phi)\sin^{2n}(\phi)\sin^{2(m+n)}(\theta)\big]\\\left[\frac{v(x-\xi)^{2(m+n)}+(1-v)(x+\xi)^{2(m+n)}}{m!n!}\right]\\
    =f_{nm}(x,v,\xi)c_{nm}(\phi,\theta).
\end{multline}
From this follows that the contribution to FI from mode $(n,m)$ is approximately equal to:
\begin{align}
\begin{split}\label{app:avg_angl}
&w^2F_{nm}(d,v,\xi)\\
\approx&\frac{\Big( \int d\phi\int d \theta\, \textbf{p}(\phi,\theta)c_{nm}(\phi,\theta)\Big)^2\Big(\frac{d}{dx}f_{nm}(x,\theta,v,\xi)\Big)^2}{4\Big(\int d\phi\int d \theta\, \textbf{p}(\phi,\theta)c_{nm}(\phi,\theta)\Big) f_{nm}(x,\theta,v,\xi)}\\\approx&\int d\phi\int d \theta\, \textbf{p}(\phi,\theta)c_{nm}(\phi,\theta)\frac{(\frac{d}{dx}f_{nm}(x,\theta,v,\xi))^2}{4f_{nm}(x,\theta,v,\xi)} \\
\approx&w^2\int d\phi\int d \theta\, \textbf{p}(\phi,\theta)F_{nm}(d,\phi,\theta,v,\xi).
\end{split}
\end{align}
Here, in order to obtain the last line, we multiply the nominator and denominator by $c_{nm}(\phi,\theta)$. From that and (\ref{eq:FI_definition}) follows that $F(d,v,\xi)$ is approximately a simple weighted average over angles $\phi$ and $\theta$:
\begin{equation}
    F(d,v,\xi)\approx\int_0^{2\pi}d\phi\int_0^{\pi}d\theta \,\textbf{p}(\phi,\theta)F(d,\phi,\theta,v,\xi).
\end{equation}
Let us make the observation that whenever $\textbf{p}(\phi,\theta)=\textbf{p}(\phi)\delta(\theta-\theta(0))$ for any value of $\theta(0)$ equality in the above formula is strict for any $x$. The reason for this is that part of  $p(n,m|x,\phi,\theta,v,\xi)$ depenendent on $\phi$ can be always factored out as in our approximate discussion of $c_{nm}(\phi,\theta)$ (see (\ref{eq:beta_nm}), and (\ref{eq:pr_nm})) . 

Explicitly calculating FI for small separations $x\ll1$ based on the leading orders of $x$ 
one finds that the terms in FI originating from modes $(n,m)$ with $n+m=1$  are constant with respect to separation, while the terms for other modes are of the order at least $O(x^2)$. The constant terms are given by:
\begin{equation}\label{eq:fnm}
   w^2F_{nm}(d|\rho')\approx \frac{C_{nm}(1-\kappa+2\kappa v)^2}{(\kappa-1)^2+4 \kappa v},
\end{equation}
where only dependence on $n,m$ is in  $C_{nm}$ which is some constant non-negative coefficient determined by the probability distribution $\textbf{p}(\phi,\theta)$. These coefficients are given by:
    \begin{multline}
        C_{nm}=\int d\phi\int d\theta\,\textbf{p}(\phi,\theta)c_{nm}(\phi,\theta)\\ \int d\phi\int d\theta\,\textbf{p}(\phi,\theta)  \sin^{2m+2n} (\theta )\sin ^{2 m}(\phi ) \cos ^{2 n}(\phi ) 
    \end{multline}
 From this we notice that summing these terms removes the dependence on $\phi$ as:
 \begin{multline}
 C:=C_{10}+C_{01}=\\\int d\phi\int d\theta\,\textbf{p}(\phi,\theta)  \sin^{2} (\theta )\Big[\sin ^{2 }(\phi )+ \cos ^{2 }(\phi )\Big] \\
 =\int d\theta\,\textbf{p}(\theta)  \sin^{2} (\theta ).
 \end{multline}
The coefficient $C$ is positive, as long as $\textbf{p}(\phi,\theta)$ is not fully concentrated on the poles, as in this case $C_{nm}=0$. Therefore, excluding those special cases where the problem of two sources is not well stated, one has FI that does not drop to 0 for $x\rightarrow0$, as the rest of the expression (\ref{eq:fnm}) is nonzero for relevant values of $\kappa$ as discussed in the main text.  This shows that the method based on the measurement in Hermite-Gauss basis is free from the Rayleigh's curse for any dynamic changes in orientation angles for the considered setup.

\section{Fisher information for direct imaging}\label{app:direct}
\setcounter{equation}{0}

 This section describes how the FI is calculated for perfect direct imaging measurements. For such a case, the FI in the stationary scenario is given by:

\begin{equation} \label{eq:FI_DI}
    F_{\textnormal{D}}(d,q_i,v,\xi) = \int_{\mathbf{R}^2} d\vec{r}\, \frac{1}{p(\vec{r}\,|q_i,d,v,\xi)} 
        \left( \frac{\partial}{\partial d} p(\vec{r}\,|q_i,d,v,\xi) \right)^2,
\end{equation}
where $p(\vec{r}\,|q_i,d,v,\xi)$ represents the probability density of measuring the photon in the location described by $\vec{r}$ and the locations of the sources are described by $q_i$ with $i=1,2$. This probability reads \cite{Linowski}:
\begin{align} \label{eq:pdi}
    p(\vec{r}\,|q_i,d,\xi)= v|u_{1,00}|^2 + (1-v)|u_{2,00}|^2,
\end{align}
where $u_{i, 00}\coloneqq u_{00}(\vec{r}-\vec{r}_i)$. In the stationary scenario $\xi$ plays no role as it only shifts the center of the probability distribution, which does not play any role in the integral (\ref{eq:FI_DI}). One can find that in the limit of $d\rightarrow 0$ FI  for $v\neq1/2$  is approximately given by:
\begin{equation}
    w^2F_D(d,q_i,v,\xi)=(2v-1)^2,
\end{equation}
and $w^2F_D(d,q_i,1/2,\xi)\approx 8x^2$ for $v=1/2$. Note that for perfect direct imaging the dependence on relative brightness $v$ is present, and when this parameter is known Rayleigh's curse is not present for unequal brightness of the sources. For dynamic scenarios, we replace the probabilities (\ref{eq:pdi}) with:
\begin{equation}
        p(\vec{r}\,|d,v,\xi)=\\ \int_\Gamma dq_i\,\textbf{p}(q_i)  p(\vec{r}\,|q_i,d,v,\xi),\label{eq:pDI}
\end{equation}
or appropriate time average. To compute approximate FI for perfect direct imaging in the regime where $x\ll1$ we proceed analogously to the case of measurement in Hermite-Gauss modes. Observe that if one has a distribution of an orientation angle $\textbf{p}(\phi)$ that has translational symmetry $\textbf{p}(\phi)=\textbf{p}(\phi+\pi)$ which is, for example, the case for rotations, and if the rotation axis is placed in the center between the sources ($\xi=0$), then the dependence on $v$ disappears as in (\ref{eq:pDI}). This is because in (\ref{eq:pDI}) one always has some integral over terms:
\begin{multline}
    \textbf{p}(\phi)\left(v|u_{1,00}(\phi)|^2 + (1-v)|u_{2,00}(\phi)|^2\right)\\+\textbf{p}(\phi+\pi)\left(v|u_{1,00}(\phi+\pi)|^2+ (1-v)|u_{2,00}(\phi+\pi)|^2\right).
\end{multline}
However, $u_{1(2),00}(\phi+\pi)=u_{(2)1,00}(\phi)$ and therefore this term results in:
\begin{equation}
    \textbf{p}(\phi)(|u_{1,00}(\phi)|^2 + |u_{2,00}(\phi)|^2),
\end{equation}
where the dependence on $v$ was removed. Therefore, for such scenarios, we have FI rather resembling the case of equal brightness, and thus with Rayleigh's curse.  Note that this is because for such distributions sources simply interchange their position effectively averaging their brightnesses. Therefore, in scenarios in which such balanced interchanges occur, the dependence on brightness will disappear. So, this is also the case when $\phi$ is independent of $\theta$ and one has symmetric distribution $\textbf{p}(\theta)$ such that $\textbf{p}(\pi/2-\psi)=\textbf{p}(\pi/2+\psi)$. 
 
\end{document}